\documentclass[pra,twocolumn,amsmath,amssymb,superscriptaddress]{revtex4-1}
\usepackage{epsfig,amsmath}
\usepackage{subfigure}
\usepackage{graphicx}
\usepackage{dcolumn}
\usepackage{stmaryrd}
\usepackage{mathrsfs}
\usepackage{pifont}
\usepackage{amsthm}
\usepackage{amssymb}
\usepackage{bm}
\usepackage{latexsym}
\usepackage[colorlinks=true,linkcolor=blue,citecolor=blue]{hyperref}
\usepackage{color}
\usepackage{epstopdf}

\usepackage{booktabs}
\usepackage{threeparttable}
\usepackage{multirow}

\begin{document}

\title{Entangling distant systems via universal nonadiabatic passage}

\author{Zhu-yao Jin}
\affiliation{School of Physics, Zhejiang University, Hangzhou 310027, Zhejiang, China}

\author{Jun Jing}
\email{Contact author: jingjun@zju.edu.cn}
\affiliation{School of Physics, Zhejiang University, Hangzhou 310027, Zhejiang, China}

\date{\today}

\begin{abstract}
In this paper, we derive universal nonadiabatic passages in a general $M+N$-dimensional discrete system, where $M$ and $N$ denote the degrees of freedom for the assistant and working subspaces, respectively, that could be separated by rotation or energy and coupled through driving. A systematic method is provided to construct parametric ancillary bases by the von Neumann equation with the time-dependent system Hamiltonian. The resulting universal passages set up connections between arbitrary initial and target states. In applications, a transitionless dynamics can be formulated to entangle distant qubits, as a crucial prerequisite for practical quantum networks. Using tunable longitudinal interaction between distant qubits and driving frequency, the superconducting qubits can be prepared from the ground state to the single-excitation Bell state with a fidelity as high as $\mathcal{F}=0.997$ and be further converted to the double-excitation Bell state with $\mathcal{F}=0.982$. Moreover, our protocol is extended to generate the Greenberger-Horne-Zeilinger state for an $N$-qubit system with $N$ steps. Our work develops a full-fledged theory for nonadiabatic state engineering, which is flexible in target selection and robust against both external noises and systematic errors in quantum information processing.
\end{abstract}

\maketitle

\section{Introduction}

Quantum entanglement~\cite{Einstein1935Can,Bohr1935Can,Horodecki2009Quantum} is an essential resource~\cite{Bennett1998Quantum} for quantum information processing~\cite{Ladd2010Quantum,Buluta2011Netureal,Kimble2008Quantum,Gisin2007Quantum,
Kimble2008Quantum,Wendin2017Quantum}. Remote entangled systems are desired for quantum communication protocols, including quantum teleportation~\cite{Bennett1993Teleporting}, quantum key distribution~\cite{Ekert1991Quantum}, quantum repeaters~\cite{Briegel1998Quantum}, quantum secret sharing~\cite{Hillery1999Quantum}, quantum secure direct communication~\cite{Long2002Theoretically}, and quantum secure communication~\cite{Deng2003Twostep}. Bell states~\cite{Brunner2014Bell} are the most popular states with maximal entanglement. They can be categorized into single-excitation and double-excitation entangled states, i.e., $(|eg\rangle\pm|ge\rangle)/\sqrt{2}$ and $(|ee\rangle\pm|gg\rangle)/\sqrt{2}$, where $|e\rangle$ and $|g\rangle$ denote the excited and ground states of the two-state systems, respectively. More generally, the maximally entangled states for $N$ particles include the Greenberger-Horne-Zeilinger (GHZ) states~\cite{Bouwmeester1999Observation,Wei2006Generation}, i.e., $(|e\rangle^{\otimes N}+|g\rangle^{\otimes N})/\sqrt{2}$, and the Werner state~\cite{Dur2000Three,Lee2000Entanglement}, i.e., $(|egg\cdots g\rangle+|geg\cdots g\rangle+\cdots+|ggg\cdots e\rangle)/\sqrt{N}$. While the latter is more robust against external noise~\cite{Brunner2014Bell} than the former; the former has a dramatic application in high-precision metrology~\cite{Bollinger1996Optimal,Vittorio2004Quantum}.

Creating the maximal entanglement between neighboring two-state systems has been proposed and realized in various platforms, including cavity QED systems~\cite{Zheng2001Onestep,Guo2002Scheme}, hybrid cavity-magnon systems~\cite{Yuan2020Steady,Qi2022Generation}, superconducting circuits~\cite{Stojanovi2020Bare,Feng2022Generation}, and cold neutral atoms~\cite{Stojanovi2021Scalable}. As a key ingredient for quantum networks~\cite{Ribordy2000Long,Hu2021Long,Zou2022Bell}, long-range entanglement has been proposed in a superconducting waveguide QED system~\cite{Zhang2023Generating}. It remains, however, a challenge to effectively engineer a high-fidelity entanglement for remote qubits due to the fast-descending coupling strength with distance. Recently, the so-called Andreev spin qubit (ASQ)~\cite{Hays2021Coherent,Pita2023Direct,Tosi2019Spin,Hays2020Continuous,
Wesdorp2023Dynamical,Wesdorp2024Microwave,Bargerbos2023Spectroscopy,Pita2024Strong}, i.e., the semiconducting spin qubit embedded into a Josephson junction~\cite{Loss1998Quantum,Hanson2007Spins}, has attracted considerable attention due to its long lifetime ($\sim20\mu$s) and small size ($\sim100$nm~\cite{Burkard2023Semiconductor}), which is of interest for large-scale quantum devices. More than the strong exchange coupling between ASQs and superconducting circuits~\cite{Hays2021Coherent,Pita2023Direct}, a strong, long-range, and tunable longitudinal interaction ($2\pi\times 0\sim200$ MHz) among ASQs~\cite{Pita2024Strong} has been experimentally established by the spin-dependent supercurrent~\cite{Hays2021Coherent,Pita2023Direct,Tosi2019Spin,Hays2020Continuous,
Wesdorp2023Dynamical,Wesdorp2024Microwave,Bargerbos2023Spectroscopy}. It could become a promising platform for generation and conversion of the maximally entangled states of distant qubits.

Quantum state engineering is primarily motivated to steer the system from a fiducial state to a predesigned target state. Adiabatic protocols, such as stimulated Raman adiabatic passage~\cite{Vitanov2017Stimulated,Guery2019Shortcuts}, have found wide-range applications due to their simplicity and error resilience. However, the long evolution time of open quantum systems~\cite{Jing2016Eigenstate} will lead to the inconvenient decoherence effect on engineering protocols such as generation of the entangled states~\cite{Marr2003Entangled,Chen2007Generation}. Various nonadiabatic methods have been proposed to accelerate the passage, such as the holonomic quantum transformation~\cite{Sjoqvist2012Nonadiabatic,Liu2019Plug,Setiawan2021Analytic,
Feng2013Experimental,Xu2022Realizing,Ming2022Experimental,Jin2024Geometric} and the shortcut to adiabaticity~\cite{Chen2010Shortcut,Guery2019Shortcuts,Baksic2016Speeding,Li2016Shortcut,
Chen2010Fast,Chen2011Lewis,Qi2022Accelerated,An2016Shortcuts}. The latter includes the Lewis-Riesenfeld theory for invariant~\cite{Chen2010Fast,Chen2011Lewis} and the counterdiabatic driving method in the adiabatic~\cite{Chen2010Shortcut} and dressed-state bases~\cite{Baksic2016Speeding,Li2016Shortcut}. They can be unified~\cite{Jin2024Shortcut} such that any nonadiabatic control can be performed along the instantaneous basis states, which satisfy the von Neumann equation with the time-dependent system Hamiltonian. Despite the versatility of this universal theory, it lacks a systematic formulation about the ancillary basis states for discrete systems of any size or multiparticle systems.

This paper presents the relation between the degrees of freedom of the time-dependent Hamiltonian and the nonadiabatic passages and proposes to generate and convert entangled states of distant qubits with a high fidelity. We separate the full Hilbert space of a general discrete system of a finite size into the working subspace and the assistant subspace, which are coupled with parametric driving and can be distinguished by, but not limited to, the energy splitting. In addition to find the nonadiabatic passages across the two subspaces for fast state engineering, we obtain the condition for recasting static dark states to useful passages. Then, we apply our theory to generate distant entanglement in a superconducting system that distant ASQs are indirectly coupled with a strong and tunable longitudinal interaction~\cite{Pita2024Strong}. Our protocol is superior to the existences in a superconducting waveguide QED system~\cite{Zhang2023Generating} with respect to the fidelity and the range of the target entangled states. We can faithfully realize the mutual conversion between the single-excitation and double-excitation Bell states. Moreover, our protocol can generate GHZ states for $N$ qubits in $N$ steps. Each step is characterized with a fixed setting for qubits and driving pulses.

The rest of this paper is structured as follows. In Sec.~\ref{Control}, we briefly review the control theory~\cite{Jin2024Shortcut} in which the transitionless dynamics is running in the ancillary picture. In Sec.~\ref{GenDisc}, we formulate the ancillary bases for a general $M+N$ system, and find a sufficient condition to activate the dark states to the nonadiabatic passages. In Sec.~\ref{illustrative}, the $1+N$ and $2+N$ systems are exemplified to further illustrate our full-fledged nonadiabatic control theory. We present a brief comparison with existing methods about pulse design and Hamiltonian engineering. Section~\ref{conversion} contributes to generating high-fidelity entangled states of distant Andreev spin qubits in superconducting systems, including Bell states and multiparticle GHZ state. We also demonstrate the conversion between orthogonal Bell states. The whole paper is concluded in Sec.~\ref{conclusion}. In the Appendix, an iterative recipe is introduced to construct the ancillary basis states for a finite-size time-dependent system.

\section{General framework}\label{general}

\subsection{Nonadiabatic control theory}\label{Control}

Consider a closed quantum system of arbitrary $K$ dimensions under a time-dependent Hamiltonian $H(t)$. The system dynamics can be described by the time-dependent Schr\"odinger equation ($\hbar\equiv1$)
\begin{equation}\label{Sch}
i\frac{d|\psi_m(t)\rangle}{dt}=H(t)|\psi_m(t)\rangle,
\end{equation}
where $|\psi_m(t)\rangle$'s are the pure-state solutions that constitute a completed set for the system space. Alternatively, the system dynamics could be described in an instantaneous picture with the orthonormal bases $|\mu_k(t)\rangle$, which span the same Hilbert space as $|\psi_m(t)\rangle$. Different from $|\psi_m(t)\rangle$, $|\mu_k(t)\rangle$ can be formally determined with no knowledge of the system Hamiltonian $H(t)$. A typical set of ancillary bases $|\mu_k(t)\rangle$ is shown in Eqs.~(\ref{AncibaseGeneral_high})-(\ref{AncibaseGeneral_across}) and an iterative recipe for constructing $|\mu_k(t)\rangle$ can be found in the Appendix. The unitary transformation linking the solution picture $|\psi_m(t)\rangle$ and the ancillary picture $|\mu_k(t)\rangle$ is shown in
\begin{equation}\label{transformation}
|\psi_m(t)\rangle=\sum_{k=1}^Kc_{mk}(t)|\mu_k(t)\rangle,
\end{equation}
where $c_{mk}(t)$ is the element of a $K\times K$ transformation matrix $\mathcal{C}$ at the $m$th row and the $k$th column. Substituting Eq.~(\ref{transformation}) into Eq.~(\ref{Sch}), we have $K^2$ differential equations in all for the matrix elements~\cite{Liu2019Plug,Jin2024Shortcut}, i.e.,
\begin{equation}\label{element}
\frac{d}{dt}c_{mk}(t)=i\sum_{n=1}^K\left[\mathcal{G}_{kn}(t)-\mathcal{D}_{kn}(t)\right]c_{mn}(t),
\end{equation}
where $\mathcal{G}_{kn}(t)\equiv i\langle\mu_k(t)|\dot{\mu}_n(t)\rangle$ and $\mathcal{D}_{kn}(t)\equiv\langle\mu_k(t)|H(t)|\mu_n(t)\rangle$ represent the geometric and dynamical parts of the proportional factor, respectively. Since each $c_{mk}(t)$ is coupled to the other $K-1$ elements $c_{mn}(t)$, $1\leq n\neq k\leq K$, it is hard to directly solve Eq.~(\ref{element}).

Dealing with the system Hamiltonian in a time-independent ancillary picture $|\mu_k(0)\rangle$ is useful to simplify and then at least partially solve Eq.~(\ref{element}). Under the unitary rotation by $V(t)\equiv\sum_{k=1}^K|\mu_k(t)\rangle\langle\mu_k(0)|$, we have
\begin{equation}\label{Hamrot}
\begin{aligned}
&H_{\rm rot}(t)=V^\dagger(t)H(t)V(t)-iV^\dagger(t)\frac{d}{dt}V(t)\\
&=-\sum_{k=1}^K\sum_{n=1}^K\left[\mathcal{G}_{kn}(t)-\mathcal{D}_{kn}(t)\right]
|\mu_k(0)\rangle\langle\mu_n(0)|.
\end{aligned}
\end{equation}
{\em If} $H_{\rm rot}(t)$ could be diagonalized in $|\mu_k(0)\rangle$ with certain $k$'s, $k\in\{1,2,\cdots,K\}$, then the relevant proportional factor $\mathcal{G}_{kn}(t)-\mathcal{D}_{kn}(t)$ in Eq.~(\ref{element}) becomes vanishing unless $n=k$ and vice versa. The diagonalization under $V(t)$ could be full or partial, which is implicitly subject to the degrees of freedom of the system Hamiltonian $H(t)$. Practically, we have proved a sufficient and necessary condition that if the projection operator $\Pi_k(t)\equiv|\mu_k(t)\rangle\langle\mu_k(t)|$ satisfies the von Neumann equation~\cite{Jin2024Shortcut}
\begin{equation}\label{von}
\frac{d}{dt}\Pi_k(t)=-i\left[H(t), \Pi_k(t)\right],
\end{equation}
then $H_{\rm rot}(t)$ is diagonalized in the basis state $|\mu_k(0)\rangle$ with a nonvanishing occupation.

For eligible $k$'s, Eq.~(\ref{element}) can be reduced to
\begin{equation}\label{digelement}
\frac{d}{dt}c_{mk}(t)=i\left[\mathcal{G}_{kk}(t)-\mathcal{D}_{kk}(t)\right]c_{mk}(t).
\end{equation}
One can immediately find the solution:
\begin{equation}\label{cmk}
\begin{aligned}
c_{mk}(t)&=e^{if_k(t)}c_{mk}(0), \\
f_k(t)&\equiv\int_0^t\left[\mathcal{G}_{kk}(t')-\mathcal{D}_{kk}(t')\right]dt'.
\end{aligned}
\end{equation}
When $\{k\}=\{1,2,\cdots,K\}$, $H_{\rm rot}(t)$ is fully diagonalized, i.e., all the projection operators can satisfy the von Neumann equation~(\ref{von}) under a common parametric setting. Consequently, Eqs.~(\ref{Sch}), (\ref{element}), and (\ref{Hamrot}) give rise to a full-rank evolution operator for the system:
\begin{equation}\label{evolve}
U(t,0)=\sum_{k=1}^{K}e^{if_k(t)}|\mu_k(t)\rangle\langle\mu_k(0)|
\end{equation}
with the generated phase $f_k(t)$ in Eq.~(\ref{cmk}). In this ideal case, the system can be steered along any ancillary basis state, $|\mu_k(0)\rangle\rightarrow|\mu_k(t)\rangle$, with no transition to others during the whole evolution.

A trivial solution to the von Neumann equation~(\ref{von}) is the time-independent dark state, i.e., $|\mu_k(t)\rangle\Rightarrow|\mu_k\rangle$, leading to a null result on both sides of Eq.~(\ref{von}) due to $H(t)|\mu_k\rangle=0$. If $|\mu_k(t)\rangle$'s are empowered with a slowly varying (adiabatic) time dependence, i.e., $d\Pi_k(t)/dt\approx0$, these seemingly useless states can support the stimulated Raman adiabatic passage~\cite{Vitanov2017Stimulated,Guery2019Shortcuts}, that could be regarded as an approximate solution to Eq.~(\ref{von}). In Eq.~(\ref{Hamrot}), the time-dependent $|\mu_k(t)\rangle$ and the to-be-diagonalized Hamiltonian $H_{\rm rot}(t)$ are mutually dependent on each other. The implicit connection between the system Hamiltonian $H(t)$ and the basis state $|\mu_k(t)\rangle$ in the von Neumann equation~(\ref{von}) remains obscure until $|\mu_k(t)\rangle$ is explicitly determined. Then we are motivated to provide an explicit formulation about these completed and orthonormal ancillary bases.

\subsection{General two-subspace system}\label{GenDisc}

\begin{figure}[htbp]
\centering
\includegraphics[width=0.9\linewidth]{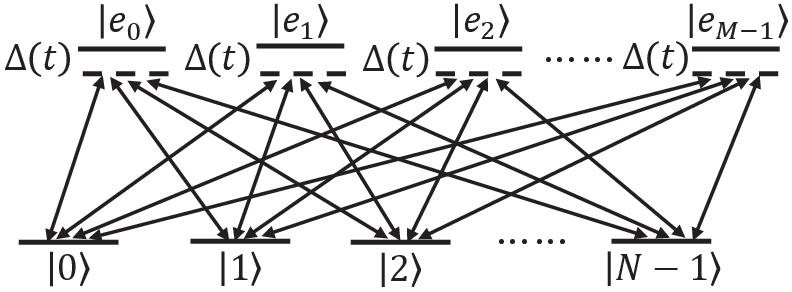}
\caption{Sketch of a general $M+N$-dimensional system under control. The transition $|n\rangle\leftrightarrow|e_m\rangle$, $0\le n\le N-1$ and $0\le m\le M-1$, is driven by the field with the Rabi frequency $\Omega_n^{(m)}(t)$ and the time-dependent phase $\varphi_n^{(m)}(t)$. The upper (assistant) and down (working) subspaces are splitted with the detuning $\Delta(t)$ in the rotating frame.}\label{generalmodel}
\end{figure}

We formulate $|\mu_k(t)\rangle$ in an arbitrary discrete model of $M+N$ levels, where $M$ and $N$ respectively denote the dimensions of the assistant and working subspaces as shown in Fig.~\ref{generalmodel}. Despite the fact that this model is treated in a degenerate situation (it is merely for finding compact expressions and yet does not lose generality of our theory), it is still a generalization for the three-level system~\cite{Vitanov2017Stimulated,Guery2019Shortcuts,Chen2010Shortcut,Guery2019Shortcuts,
Baksic2016Speeding,Li2016Shortcut,Chen2010Fast,Chen2011Lewis,An2016Shortcuts,
Sjoqvist2012Nonadiabatic,Liu2019Plug}, the four-level tripod system~\cite{Vitanov2017Stimulated,Setiawan2021Analytic}, and other multilevel systems~\cite{Vitanov2017Stimulated}. These pedagogical models are popular in the stimulated Raman adiabatic passage~\cite{Vitanov2017Stimulated,Guery2019Shortcuts}, the shortcut to adiabaticity~\cite{Chen2010Shortcut,Guery2019Shortcuts,Baksic2016Speeding,Li2016Shortcut,Chen2010Fast,Chen2011Lewis,
Qi2022Accelerated,An2016Shortcuts}, the adiabatic geometric gate~\cite{Jones2000Geometric,Duan2001Geometric}, and the nonadiabatic holonomic quantum transformation~\cite{Sjoqvist2012Nonadiabatic,Liu2019Plug,Setiawan2021Analytic}. The current model contains all the necessary aspects of the problem at hand and sets our investigation on a clear motivation.

We focus on the driving-based state engineering across the assistant and working subspaces. The transition between the upper level $|e_m\rangle$, $m=0, 1, \cdots, M-1$, and the down level $|n\rangle$, $n=0, 1, \cdots, N-1$, is induced by the driving field with the time-dependent Rabi frequency $\Omega_n^{(m)}(t)$, phase $\varphi_n^{(m)}(t)$, and detuning $\Delta(t)$. Then, the full Hamiltonian can be written as
\begin{equation}\label{Hamgeneral}
\begin{aligned}
&H(t)=\Delta(t)\sum_{m=0}^{M-1}|e_m\rangle\langle e_m|\\
&+\left[\sum_{m=0}^{M-1}\sum_{n=0}^{N-1}\Omega_n^{(m)}(t)e^{i\varphi_n^{(m)}(t)}|e_m\rangle\langle n|+{\rm H.c.}\right].
\end{aligned}
\end{equation}
It is clearly not the most general configuration due to the lack of the internal driving fields in both subspaces, and then it cannot afford a full-rank nonadiabatic control over the whole system. Nevertheless, it is sufficient to have at least two nonadiabatic paths for state transfer across the two subspaces as long as all $\Omega_n^{(m)}(t)$ are nonvanishing and tunable in time.

In the theoretical framework for universal control (see Sec.~\ref{Control}), the system dynamics can be faithfully described in an $M+N$-dimensional ancillary picture. Using the iterative recipe in the Appendix, one can formulate a completed and orthonormal set of ancillary bases. In the assistant subspace, we have
\begin{equation}\label{AncibaseGeneral_high}
\begin{aligned}
|\tilde{\mu}_0(t)\rangle&=\sin\tilde{\theta}_0(t)|e_0\rangle+\cos\tilde{\theta}_0(t)e^{-i\tilde{\alpha}_0(t)}|e_1\rangle,\\
|\tilde{\mu}_1(t)\rangle&=\sin\tilde{\theta}_1(t)|\tilde{b}_0(t)\rangle+\cos\tilde{\theta}_1(t)e^{-i\tilde{\alpha}_1(t)}|e_2\rangle,\\
&\cdots\\
|\tilde{\mu}_{M-2}(t)\rangle&=\sin\tilde{\theta}_{M-2}(t)|\tilde{b}_{M-3}(t)\rangle\\
&+\cos\tilde{\theta}_{M-2}(t)e^{-i\tilde{\alpha}_{M-2}(t)}|e_{M-1}\rangle.
\end{aligned}
\end{equation}
In the working subspace, we have
\begin{equation}\label{AncibaseGeneral_low}
\begin{aligned}
|\mu_0(t)\rangle&=\cos\theta_0(t)|0\rangle-\sin\theta_0(t)e^{-i\alpha_0(t)}|1\rangle,\\
|\mu_1(t)\rangle&=\cos\theta_1(t)|b_0(t)\rangle-\sin\theta_1(t)e^{-i\alpha_1(t)}|2\rangle,\\
&\cdots\\
|\mu_{N-2}(t)\rangle&=\cos\theta_{N-2}(t)|b_{N-3}(t)\rangle \\ &-\sin\theta_{N-2}(t)e^{-i\alpha_{N-2}(t)}|N-1\rangle.
\end{aligned}
\end{equation}
And, the rest bases are across the two subspaces:
\begin{equation}\label{AncibaseGeneral_across}
\begin{aligned}
|\mu_{N-1}(t)\rangle&=\cos\phi(t)|b_{N-2}(t)\rangle\\ &-\sin\phi(t)e^{-i\alpha(t)}|\tilde{b}_{M-2}(t)\rangle,\\
|\mu_N(t)\rangle&=\sin\phi(t)|b_{N-2}(t)\rangle+\cos\phi(t)^{-i\alpha(t)}|\tilde{b}_{M-2}(t)\rangle.
\end{aligned}
\end{equation}
Here, the undetermined parameters $\tilde{\theta}_m(t)$, $\theta_n(t)$, and $\phi(t)$ are in charge of the population transfer along the nonadiabatic passage; $\tilde{\alpha}_m(t)$, $\alpha_n(t)$, and $\alpha(t)$ control the relative phase among the discrete states; and the $M+N-2$ bright states are defined as
\begin{equation}\label{brightGeneral}
\begin{aligned}
|\tilde{b}_0(t)\rangle&\equiv\cos\tilde{\theta}_0(t)|e_0\rangle-\sin\tilde{\theta}_0(t)e^{-i\tilde{\alpha}_0(t)}|e_1\rangle,\\
|\tilde{b}_1(t)\rangle&\equiv\cos\tilde{\theta}_1(t)|\tilde{b}_0(t)\rangle
-\sin\tilde{\theta}_1(t)e^{-i\tilde{\alpha}_1(t)}|e_2\rangle,\\
&\cdots\\
|\tilde{b}_{M-2}(t)\rangle&\equiv\cos\tilde{\theta}_{M-2}(t)|\tilde{b}_{M-3}(t)\rangle\\
&-\sin\tilde{\theta}_{M-2}(t)e^{-i\tilde{\alpha}_{M-2}(t)}|e_{M-1}\rangle,\\
|b_0(t)\rangle&\equiv\sin\theta_0(t)|0\rangle+\cos\theta_0(t)e^{-i\alpha_0(t)}|1\rangle,\\
|b_1(t)\rangle&\equiv\sin\theta_1(t)|b_0(t)\rangle+\cos\theta_1(t)e^{-i\alpha_1(t)}|2\rangle,\\
&\cdots\\
|b_{N-2}(t)\rangle&\equiv\sin\theta_{N-2}(t)|b_{N-3}(t)\rangle\\ &+\cos\theta_{N-2}(t)e^{-i\alpha_{N-2}(t)}|N-1\rangle.
\end{aligned}
\end{equation}
In the same subspace, the bright states and the ancillary bases with the same subscript are orthonormal to each other, i.e., $\langle\tilde{\mu}_m(t)|\tilde{b}_m(t)\rangle=0$ and $\langle\mu_n(t)|b_n(t)\rangle=0$.

The two ancillary bases in Eq.~(\ref{AncibaseGeneral_across}), which support the universal control, also satisfy $\langle\mu_{N-1}(t)|\mu_N(t)\rangle=0$. By substituting them into the von Neumann equation~(\ref{von}) with the Hamiltonian $H(t)$ in Eq.~(\ref{Hamgeneral}), the phases, the Rabi frequencies, and the detuning are found to be
\begin{equation}\label{ConditionEff}
\begin{aligned}
&\varphi_n^{(m)}(t)=\varphi(t)-\tilde{\alpha}_{m-1}(t)+\alpha_{n-1}(t),  \\
&\Omega_n^{(m)}(t)=-\Omega(t)\sin\tilde{\theta}_{m-1}(t)\prod_{m'=m}^{M-2}\cos\tilde{\theta}_{m'}(t)\\
&\times\cos\theta_{n-1}(t)\prod_{n'=n}^{N-2}\sin\theta_{n'}(t), \\
&\Delta(t)=\dot{\alpha}(t)-2\dot{\phi}(t)\cot\left[\varphi(t)+\alpha(t)\right]\cos2\phi(t),\\
&\Omega(t)=-\dot{\phi}(t)/\sin[\varphi(t)+\alpha(t)],
\end{aligned}
\end{equation}
where $n$ ($m$) runs from $0$ to $N-1$ ($M-1$), $\tilde{\alpha}_{-1}(t)\equiv0$, $\alpha_{-1}(t)\equiv0$, $\sin\tilde{\theta}_{-1}(t)\equiv-1$, and $\cos\theta_{-1}(t)\equiv1$. Note that $\Omega(t)$ scales the driving intensity and $\varphi(t)$ is an arbitrary function of time with no singularity.

We then substitute the ancillary bases of Eqs.~(\ref{AncibaseGeneral_high}) and (\ref{AncibaseGeneral_low}) into the von Neumann equation~(\ref{von}). Under those conditions of Eq.~(\ref{ConditionEff}), the relevant parameters have to be time independent, i.e.,
\begin{equation}\label{parameterEff}
\begin{aligned}
&\tilde{\theta}_m(t)\rightarrow\tilde{\theta}_m, \quad \tilde{\alpha}_m(t)\rightarrow\tilde{\alpha}_m, \quad m=0,1,\cdots,M-2,\\
&\theta_n(t)\rightarrow\theta_n, \quad \alpha_n(t)\rightarrow\alpha_n, \quad n=0,1,\cdots,N-2.
\end{aligned}
\end{equation}
In this case, the ancillary bases in the assistant and working subspaces become static and some of them are even dark states, i.e., $|\tilde{\mu}_m(t)\rangle\rightarrow|\tilde{\mu}_m\rangle$, $|\mu_n(t)\rangle\rightarrow|\mu_n\rangle$, $H(t)|\tilde{\mu}_m\rangle=\Delta(t)|\tilde{\mu}_m\rangle$, and $H(t)|\mu_n\rangle=0$.

These results can be transparently demonstrated in the evolution operator~(\ref{evolve}), which are expressed as
\begin{equation}\label{evolveGeneral}
\begin{aligned}
&U(t,0)=\sum_{m=0}^{M-2}e^{i\tilde{f}_m(t)}|\tilde{\mu}_m\rangle\langle\tilde{\mu}_m|
+\sum_{n=0}^{N-2}e^{if_n(t)}|\mu_n\rangle\langle\mu_n|\\
&+e^{if_{N-1}(t)}|\mu_{N-1}(t)\rangle\langle\mu_{N-1}(0)|+e^{if_N(t)}|\mu_N(t)\rangle\langle\mu_N(0)|,
\end{aligned}
\end{equation}
under our Hamiltonian~(\ref{Hamgeneral}). By the definitions in Eq.~(\ref{cmk}), the generated phases are found to be
\begin{equation}
\begin{aligned}
  \tilde{f}_m(t)&=\int_0^t\left[\tilde{\mathcal{G}}_{mm}(t')-\tilde{\mathcal{D}}_{mm}(t')\right]dt', \\   f_n(t)&=0, \quad 0\leq n\leq N-2 \\
  f_{N-1}(t)&=\int_0^t\left[\mathcal{G}_{N-1,N-1}(t')-\mathcal{D}_{N-1,N-1}(t')\right]dt', \\  f_N(t)&=\int_0^t\left[\dot{\alpha}(t')-\Delta(t')\right]dt'-f_{N-1}(t),
\end{aligned}
\end{equation}
where $\tilde{\mathcal{G}}_{mm}(t')=0$, $\tilde{\mathcal{D}}_{mm}(t')=\Delta(t')$, $\mathcal{G}_{N-1,N-1}(t')=\dot{\alpha}(t')\sin^2\phi(t')$, and $\mathcal{D}_{N-1,N-1}(t')=\Delta(t')\sin^2\phi(t')-\Omega(t')\sin2\phi(t')
\cos[\varphi(t')+\alpha(t')]$. Thus, the ancillary basis state in the assistant subspace $|\tilde{\mu}_m\rangle$ might gain a global phase $\tilde{f}_m(t)$ that is purely dynamical, and that in the working subspace $|\mu_n\rangle$ is purely static. In contrast to those trivial passages, the ancillary base states $|\mu_{N-1}(t)\rangle$ and $|\mu_N(t)\rangle$ can be used to transfer population and accumulate the relative phase for the desired levels.

Alternatively, the universal passage could be observed through the system Hamiltonian $H(t)$~(\ref{Hamgeneral}) expressed with the ancillary bases in Eqs.~(\ref{AncibaseGeneral_high})-(\ref{AncibaseGeneral_across}) and the bright states in Eq.~(\ref{brightGeneral}). Under the conditions in Eqs.~(\ref{ConditionEff}) and (\ref{parameterEff}), we have
\begin{equation}\label{HamGenEff}
\begin{aligned}
&H(t)=\Delta(t)\left[\sum_{m=0}^{M-2}|\tilde{\mu}_m\rangle\langle\tilde{\mu}_m|
+|\tilde{b}_{M-2}\rangle\langle \tilde{b}_{M-2}|\right]\\
&+\left[\Omega(t)e^{i\varphi(t)}|\tilde{b}_{M-2}\rangle\langle b_{N-2}|+{\rm H.c.}\right].
\end{aligned}
\end{equation}
The first line of Eq.~(\ref{HamGenEff}) contributes to the dynamical phase $\tilde{f}_m(t)$ on the passage $|\tilde{\mu}_m\rangle$ in Eq.~(\ref{AncibaseGeneral_high}). The second line explains the Rabi oscillation between $|\tilde{b}_{M-2}\rangle$ and $|b_{N-2}\rangle$ in Eq.~(\ref{brightGeneral}), which is a sufficient condition for constructing the nonadiabatic passages along $|\mu_{N-1}(t)\rangle$ or $|\mu_N(t)\rangle$ by the von Neumann equation~(\ref{von}).

Moreover, it is found that any static base $|\tilde{\mu}_m\rangle$, $m\in\{0,1,\cdots, M-2\}$, could be recast to a full-featured state-transfer path with extra driving fields in the assistant subspace. For example, consider the revised full Hamiltonian
\begin{equation}\label{Hamrank3}
\begin{aligned}
H'(t)&=H(t)+\tilde{h}(t), \\
\tilde{h}(t)&=\delta(t)|e_{m+1}\rangle\langle e_{m+1}|\\
&+\left[\sum_{n=0}^m\omega_n^{(m+1)}(t)e^{i\Phi_n^{(m+1)}(t)}|e_{m+1}\rangle\langle e_n|+{\rm H.c.}\right],
\end{aligned}
\end{equation}
where $\delta(t)$ is the extra detuning for the level $|e_{m+1}\rangle$, $\omega_n^{(m+1)}(t)$ is the Rabi frequency of the transition $|e_{m+1}\rangle\leftrightarrow|e_n\rangle$, and $\Phi_n^{(m+1)}(t)$ is a time-dependent phase. Substituting the ancillary base $|\tilde{\mu}_m(t)\rangle$ of Eq.~(\ref{AncibaseGeneral_high}) and the ancillary bases $|\mu_{N-1}(t)\rangle$ and $|\mu_N(t)\rangle$ of Eq.~(\ref{AncibaseGeneral_across}) into the von Neumann equation (\ref{von}) with the system Hamiltonian (\ref{Hamrank3}), we can obtain
\begin{equation}\label{ConditionEffAny}
\begin{aligned}
&\Phi_n^{(m+1)}(t)=\frac{\pi}{2}-\tilde{\alpha}_m(t)+\tilde{\alpha}_{n-1}, \quad 0\leq n\leq m  \\
&\omega_n^{(m+1)}(t)=-\omega(t)\sin\tilde{\theta}_{n-1}\prod_{m'=n}^{m-1}\cos\tilde{\theta}_{m'},\\
&\omega(t)=-\dot{\theta}_m(t), \\
&\delta(t)=\dot{\tilde{\alpha}}_m(t),
\end{aligned}
\end{equation}
in addition to Eq.~(\ref{ConditionEff}) and the second line of Eq.~(\ref{parameterEff}). Equivalently, we find that the auxiliary system Hamiltonian
\begin{equation}
  \tilde{h}(t)=\omega(t)e^{i\pi/2-i\tilde{\alpha}_m(t)}|e_{m+1}\rangle\langle\tilde{b}_{m-1}|+{\rm H.c.}
\end{equation}
yields the third nonadiabatic passage $|\tilde{\mu}_m(t)\rangle$ under the conditions of Eqs.~(\ref{ConditionEff}), (\ref{parameterEff}), and (\ref{ConditionEffAny}).

Similarly, if one intends to recast a dark state $|\mu_n\rangle$, $n\in\{0,1,\cdots,N-2\}$, in the working subspace to a nonadiabatic passage, then one has to add an ancillary system Hamiltonian $h(t)$ to the original system Hamiltonian~(\ref{Hamgeneral}) as long as $\langle n+1|h(t)|b_{n-1}\rangle$ is nonvanishing and time dependent.

\subsection{Illustrative examples}\label{illustrative}

\begin{figure}[htbp]
\centering
\includegraphics[width=0.9\linewidth]{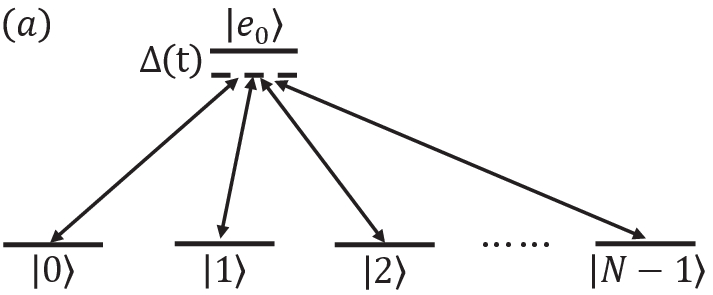}
\includegraphics[width=0.9\linewidth]{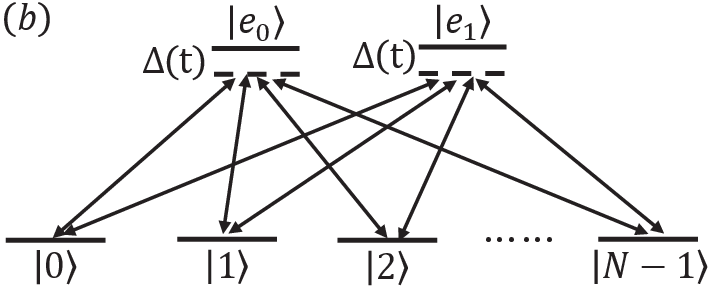}
\caption{Sketch of (a) $1+N$-dimensional and (b) $2+N$-dimensional systems under control. The transition $|n\rangle\leftrightarrow|e_m\rangle$, $0\le n\le N-1$ and $m=0, 1$, is driven by the field with Rabi frequency $\Omega_n^{(m)}(t)$ and phase $\varphi_n^{(m)}(t)$. }\label{modelN2}
\end{figure}

In this subsection, we exemplify the general model in Fig.~\ref{generalmodel} with the $1+N$-dimensional and $2+N$-dimensional systems as shown in Figs.~\ref{modelN2}(a) and \ref{modelN2}(b), respectively, to further illustrate the universal control theory in Sec.~\ref{GenDisc}. Here, the nonadiabatic passage is either $|\mu_{N-1}(t)\rangle$ or $|\mu_N(t)\rangle$.

For the $1+N$ system in Fig.~\ref{modelN2}(a), the transition between $|e_0\rangle$ and $|n\rangle$, where $n$ runs from $0$ to $N-1$, is driven by the laser field $\Omega_n^{(0)}(t)$ with the time-dependent phase $\varphi_n^{(0)}(t)$ and detuning $\Delta(t)$. Then, the full Hamiltonian can be written as
\begin{equation}\label{HamN}
\begin{aligned}
&H(t)=\Delta(t)|e_0\rangle\langle e_0|\\
+&\left[\sum_{n=0}^{N-1}\Omega_n^{(0)}(t)e^{i\varphi_n^{(0)}(t)}|e_0\rangle\langle n|+{\rm H.c.}|\right],
\end{aligned}
\end{equation}
It can be obtained from Eq.~(\ref{Hamgeneral}) when the driving field $\Omega_n^{(m)}$ with $1\leq m\leq M-2$ is turned off. In this case, the ancillary picture can be formulated by the ancillary bases for the working subspace, which are exactly the same as Eq.~(\ref{AncibaseGeneral_low}), and those mixing the working subspace and the upper level $|e_0\rangle$:
\begin{equation}\label{AncillaryN_across}
\begin{aligned}
|\mu_{N-1}(t)\rangle&=\cos\phi(t)|b_{N-2}(t)\rangle-\sin\phi(t)e^{-i\alpha(t)}|e_0\rangle,\\
|\mu_N(t)\rangle&=\sin\phi(t)|b_{N-2}(t)\rangle+\cos\phi(t)e^{-i\alpha(t)}|e_0\rangle,
\end{aligned}
\end{equation}
where the time-dependent parameter $\phi(t)$ and the bright state $|b_{N-2}(t)\rangle$ are the same as those in Eqs.~(\ref{AncibaseGeneral_across}) and (\ref{brightGeneral}), respectively.

Substituting the ancillary bases $|\mu_{N-1}(t)\rangle$ and $|\mu_N(t)\rangle$ in Eq.~(\ref{AncillaryN_across}) to the von Neumann equation~(\ref{von}) with the Hamiltonian (\ref{HamN}), we have the phases, the Rabi frequencies, and the detuning as
\begin{equation}\label{ConditionN}
\begin{aligned}
&\varphi_n^{(0)}(t)=\varphi(t)+\alpha_{n-1}(t), \quad 0\leq n\leq N-1, \\
&\Omega_n^{(0)}(t)=\Omega(t)\cos\theta_{n-1}(t)\prod_{n'=n}^{N-2}\sin\theta_{n'}(t),\\
&\Delta(t)=\dot{\alpha}(t)-2\dot{\phi}(t)\cot\left[\varphi(t)+\alpha(t)\right]\cos2\phi(t).
\end{aligned}
\end{equation}
The time-evolution operator~(\ref{evolveGeneral}) for the full system now becomes
\begin{equation}\label{evolveN1}
\begin{aligned}
&U(t,0)=\sum_{n=0}^{N-2}|\mu_n\rangle\langle\mu_n|+e^{if_{N-1}(t)}|\mu_{N-1}(t)\rangle\langle\mu_{N-1}(0)|\\
&+e^{if_N(t)}|\mu_N(t)\rangle\langle\mu_N(0)|.
\end{aligned}
\end{equation}
Along the passages $|\mu_{N-1}(t)\rangle$ and $|\mu_N(t)\rangle$, the state transfer and the accumulation of the relative phases can be demonstrated on the discrete states $|b_{N-2}(t)\rangle$ and $|e_0\rangle$.

When $N=3$, the general model in Fig.~\ref{modelN2}(a) describes a four-level tripod system~\cite{Li2018Hamiltonian,Setiawan2021Analytic}. It is found that the evolution passages in Eq.~(\ref{AncillaryN_across}) and the parametric conditions in Eq.~(\ref{ConditionN}) can be used to construct the geometric gates and population transfer among the system states.

For the $2+N$-dimensional system in Fig.~\ref{modelN2}(b), the transition between $|e_m\rangle$, $m=0, 1$, and $|n\rangle$, $n=0, 1, \cdots, N-1$, is driven by the laser field $\Omega_n^{(m)}(t)$ with the time-dependent phase $\varphi_n^{(m)}(t)$ and the same detuning $\Delta(t)$. Then the full Hamiltonian can be written as
\begin{equation}\label{HamN2}
\begin{aligned}
&H(t)=\Delta(t)(|e_0\rangle\langle e_0|+|e_1\rangle\langle e_1|) \\
&+\left[\sum_{m=0}^{1}\sum_{n=0}^{N-1}\Omega_n^{(m)}(t)e^{i\varphi_n^{(m)}(t)}
|e_m\rangle\langle n|+{\rm H.c.}\right].
\end{aligned}
\end{equation}
In comparison to the case of $M=1$ in Fig.~\ref{modelN2}(a), we now have an ancillary base in the assistant subspace, i.e., $|\tilde{\mu}_0(t)\rangle$ in Eq.~(\ref{AncibaseGeneral_high}). The ancillary bases $|\mu_n(t)\rangle$, $n<N-1$, for the work subspace remain the same as in Eq.~(\ref{AncibaseGeneral_low}). Also, the last two bases across two subspaces are the same as Eq.~(\ref{AncibaseGeneral_across}) with $M=2$. In this case, the essence of Eqs.~(\ref{AncibaseGeneral_high})-(\ref{AncibaseGeneral_across}) is equivalent to a general Morris-Shore transformation~\cite{Shkolnikov2020Effective}. When $M=2$ and $N=3$, it recovers the result in Ref.~\cite{Morris1983Reduction}.

On substituting $|\mu_{N-1}(t)\rangle$ and $|\mu_N(t)\rangle$ into the von Neumann equation~(\ref{von}) with the Hamiltonian in Eq.~(\ref{HamN2}), the time-dependent phases, Rabi frequencies, and detuning are found to be
\begin{equation}\label{ConditionN2Rabi}
\begin{aligned}
&\varphi_n^{(0)}(t)=\varphi(t)+\alpha_{n-1}(t), \quad 0\leq n\leq N-1, \\
&\varphi_n^{(1)}(t)=\varphi(t)-\left[\tilde{\alpha}_0(t)-\alpha_{n-1}(t)\right],\\
&\Omega_n^{(0)}=\Omega(t)\cos\theta_{n-1}(t)\cos\tilde{\theta}_0(t)\prod_{n'=n}^{N-2}\sin\theta_{n'}(t),\\
&\Omega_n^{(1)}=-\Omega(t)\cos\theta_{n-1}(t))\sin\tilde{\theta}_0(t)\prod_{n'=n}^{N-2}\sin\theta_{n'}(t),\\
&\Delta(t)=\dot{\alpha}(t)-2\dot{\phi}(t)\cot\left[\varphi(t)+\alpha(t)\right]\cos2\phi(t).
\end{aligned}
\end{equation}
They give exactly rise to the nonadiabatic evolution operator~(\ref{evolveGeneral}) with $M=2$.

Note our theory has provided an exact solution, [see, e.g., Eq.~(\ref{evolveGeneral})], to the time-dependent Schr\"odinger equation for a general $M+N$-dimensional system in Fig.~\ref{generalmodel}. On the contrary, it is not easy to scale up the existing methods to larger systems. For example, the pulse design method~\cite{Kang2018Pulse} expresses the system Hamiltonian with the generators of a Lie algebra and uses the Lie transform to find an evolution passage, which is now limited to the three-level and four-level (ladder type) systems. In addition, the unit fidelity is precluded by the singularity of certain experimental parameters. The Hamiltonian engineering method~\cite{Li2018Hamiltonian,Li2018qubit} focuses on the four-level systems, in which the population dynamics is presented by the rotations in the four-dimensional Euclidean space. It is a challenge to analytically treat an even larger rotation matrix~\cite{Li2018Hamiltonian}.

\section{Preparing distant qubits into maximal entangled state}\label{conversion}

\subsection{Two-qubit system}\label{Twoentangle}

\begin{figure}[htbp]
\centering
\includegraphics[width=0.9\linewidth]{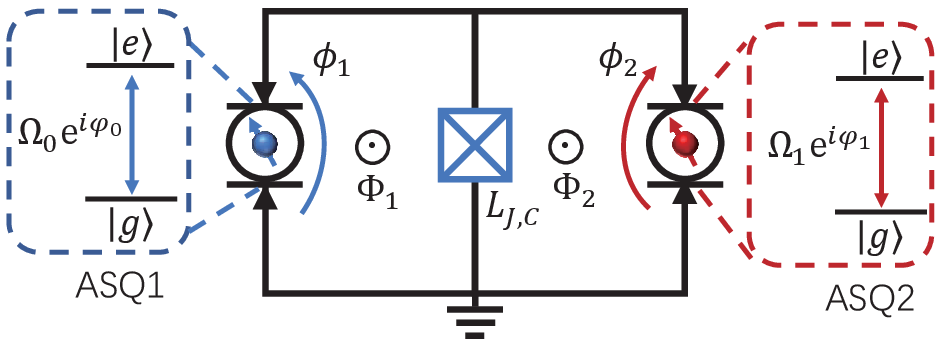}
\caption{Sketch of two longitudinally coupled Andreev spin qubits, that connect to a Josephson junction with a tunable inductance $L_{J,C}$. ASQ1 and ASQ2 are respectively driven by the laser fields with the Rabi frequencies $\Omega_0(t)$ and $\Omega_1(t)$ and the time-dependent phases $\varphi_0(t)$ and $\varphi_1(t)$. $\Phi_1$ and $\Phi_2$ are the magnetic fluxes through the two loops.}\label{modelqubit}
\end{figure}

Our application is conducted in a superconducting system~\cite{Pita2024Strong} as shown in Fig.~\ref{modelqubit}, which consists of two Andreev spin qubits~\cite{Hays2021Coherent,Pita2023Direct} and a Josephson junction and forms a double-loop superconducting quantum interference device. The full Hamiltonian reads~\cite{Pita2024Strong}
\begin{equation}\label{Hamqubit}
H=\frac{\omega_1}{2}\sigma_1^z+\frac{\omega_2}{2}\sigma_2^z+\frac{J}{2}\sigma_1^z\sigma_2^z,
\end{equation}
where $\omega_n$ and $\sigma_n^z$ denote the transition frequency and the Pauli $Z$ matrix of the $n$th qubit, respectively. For simplicity, we set $\omega_1=\omega_2=\omega$. The longitudinal interaction between two ASQs can be effectively established via the Josephson junction for the spin-dependent supercurrent of ASQ1, which induces a spin-dependent flux difference over ASQ2 and then changes its transition frequency by $J$~\cite{Pita2024Strong}. It is tunable by the Josephson inductance $L_{J,C}$, the external magnetic field, and the flux $\Phi_1$ and $\Phi_2$ through the loop. For the spins separated by a distance of $\sim 25\mu$m~\cite{Pita2024Strong}, $J/2\pi$ ranges from $0$ to $200$ MHz.

To generate entanglement in the qubits via the nonadiabatic control in Sec.~\ref{general}, we consider that the ASQ1 and ASQ2 are driven by the laser fields. Then, the full Hamiltonian can be written as
\begin{equation}\label{Hamfull}
H_{\rm tot}(t)=H+H_d(t).
\end{equation}
Here, $H$ is the original Hamiltonian in Eq.~(\ref{Hamqubit}) and $H_d(t)$ is the driving Hamiltonian:
\begin{equation}\label{HamqubDrive}
\begin{aligned}
H_d(t)&=\Omega_0(t)e^{i[\omega_0t+\varphi_0(t)]}|e\rangle_1\langle g|\\
&+\Omega_1(t)e^{i[\omega_0t+\varphi_1(t)]}|e\rangle_2\langle g|+{\rm H.c.},
\end{aligned}
\end{equation}
where $\Omega_0(t)$ and $\Omega_1(t)$ are Rabi frequencies, $\omega_0$ is the common driving frequency that is tunable at will, and $\varphi_0(t)$ and $\varphi_1(t)$ are the time-dependent phases. In the rotating frame with respect to $H$~(\ref{Hamqubit}), we have
\begin{equation}\label{HamqubRot}
\begin{aligned}
&H_{\rm rot}(t)=e^{i(\omega+\omega_0+J)t}\\
\times&[\Omega_0(t)e^{i\varphi_0(t)}(|ee\rangle\langle ge|+e^{-i2Jt}|eg\rangle\langle gg|)\\
+&\Omega_1(t)e^{i\varphi_1(t)}(|ee\rangle\langle eg|+e^{-i2Jt}|ge\rangle\langle gg|)]+{\rm H.c.}.
\end{aligned}
\end{equation}

\begin{figure}[htbp]
\centering
\includegraphics[width=0.9\linewidth]{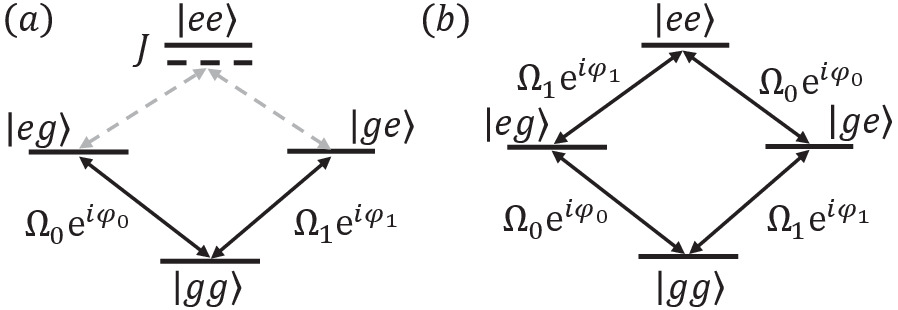}
\caption{Transition diagram for the longitudinally coupled two-ASQ system when (a) the driving frequency $\omega_0=-\omega+J$ and the coupling strength $J\gg\{\Omega_0, \Omega_1\}$, and (b) $\omega_0=-\omega$ and $J=0$.}\label{transition}
\end{figure}

\emph{Step 1}--- In this step, we intend to evolve the system from the ground state $|gg\rangle$ to the single-excitation Bell state $(|eg\rangle+|ge\rangle)/\sqrt{2}$. Using the driving frequency $\omega_0=-\omega+J$ and a strong longitudinal interaction $J\gg\{\Omega_0, \Omega_1\}$, the Hamiltonian~(\ref{HamqubRot}) becomes
\begin{equation}\label{Hamqubeff1}
H_{\rm eff}^{(1)}=\Omega_0(t)e^{i\varphi_0(t)}|eg\rangle\langle gg|+\Omega_1(t)e^{i\varphi_1(t)}|ge\rangle\langle gg|+{\rm H.c.}
\end{equation}
under the rotating-wave approximation (RWA). As shown in Fig.~\ref{transition}(a), $H_{\rm eff}^{(1)}$ describes a $1+2$-dimensional system in the subspace spanned by $|gg\rangle$, $|eg\rangle$, and $|ge\rangle$. The model has been illustrated in Sec.~\ref{illustrative}. The transitions $|ee\rangle\leftrightarrow|eg\rangle$ and $|ee\rangle\leftrightarrow|ge\rangle$ indicated by the gray dashed lines in Fig.~\ref{transition}(a) are suppressed by the strong coupling $J$.

As a special case of Eqs.~(\ref{AncibaseGeneral_high})-(\ref{AncibaseGeneral_across}) for $M=1$ and $N=2$, the ancillary bases for the working subspace can be chosen as
\begin{equation}\label{AncillaryBell_work}
|\mu_0(t)\rangle=\cos\theta_0(t)|eg\rangle-\sin\theta_0(t)e^{-i\alpha_0(t)}|ge\rangle,
\end{equation}
and those mixing the working subspace and the ground level $|gg\rangle$ are
\begin{equation}\label{AncillaryBell}
\begin{aligned}
|\mu_1(t)\rangle&=\cos\phi(t)|b_0(t)\rangle-\sin\phi(t)e^{-i\alpha(t)}|gg\rangle,\\
|\mu_2(t)\rangle&=\sin\phi(t)|b_0(t)\rangle+\cos\phi(t)e^{-i\alpha(t)}|gg\rangle,
\end{aligned}
\end{equation}
where $\alpha_0(t)$, $\alpha(t)$, $\theta_0(t)$, and $\phi(t)$ are the time-dependent parameters and the bright state is $|b_0(t)\rangle\equiv\sin\theta_0(t)|eg\rangle+\cos\theta_0(t)e^{-i\alpha_0(t)}|ge\rangle$.

Substituting $|\mu_1(t)\rangle$ and $|\mu_2(t)\rangle$ in Eq.~(\ref{AncillaryBell}) to the von Neumann equation~(\ref{von}) with the Hamiltonian~(\ref{Hamqubeff1}), we have the phases and the Rabi frequencies as
\begin{equation}\label{conditionBell}
\begin{aligned}
&\varphi_0(t)=\varphi(t), \\
&\varphi_1(t)=\varphi(t)+\alpha_0(t),\\
&\Omega_0(t)=\Omega(t)\cos\theta_0(t),\\
&\Omega_1(t)=-\Omega(t)\sin\theta_0(t),\\
&\dot{\alpha}(t)=-2\dot{\phi}(t)\cot\left[\varphi(t)+\alpha(t)\right]\cot2\phi(t).
\end{aligned}
\end{equation}

$|\mu_0(t)\rangle$ in Eq.~(\ref{AncillaryBell_work}) is a dark state. Depending on the boundary conditions of the phases $\alpha(t)$ and $\phi(t)$, the system can evolve along either $|\mu_1(t)\rangle$ or $|\mu_2(t)\rangle$. If $\alpha(t)=\pi$, $\phi(0)=\pi/2$, and $\phi(T)=0$ with $T$ the period of state transfer, then the system can evolve from the ground state $|gg\rangle$ to the bright state $|b_0(t)\rangle$ along the passage $|\mu_1(t)\rangle$. The population distribution on the bases $|eg\rangle$ and $|ge\rangle$ of the target state depends on the parameter $\theta_0(t)$ determined by the ratio of the driving intensities in Eq.~(\ref{conditionBell}). In addition, the relative phase $\alpha_0(t)$ can be tuned by the phase difference between $\varphi_0(t)$ and $\varphi_1(t)$. With no loss of generality, we set $\theta_0=\pi/4$ and $\alpha_0=0$, and then the system evolves to the single-excitation Bell state, i.e., $|gg\rangle\rightarrow(|eg\rangle+|ge\rangle)/\sqrt{2}$.

\emph{Step 2}--- In this step, we target to perform the mutual conversion between the single-excitation Bell state and the double-excitation Bell state, i.e., $(|eg\rangle+|ge\rangle)/\sqrt{2}\leftrightarrow(|ee\rangle-|gg\rangle)/\sqrt{2}$. We turn off the longitudinal interaction, i.e., $J=0$, and set the driving frequency as $\omega_0=-\omega$. The Hamiltonian~(\ref{HamqubRot}) then becomes
\begin{equation}\label{HameffGHZ}
\begin{aligned}
H_{\rm eff}^{(2)}&=\Omega_0(t)e^{i\varphi_0(t)}(|ee\rangle\langle ge|+|eg\rangle\langle gg|)\\
&+\Omega_1(t)e^{i\varphi_1(t)}(|ee\rangle\langle eg|+|ge\rangle\langle gg|)+{\rm H.c.},
\end{aligned}
\end{equation}
which forms a $2+2$-dimensional system as shown in Fig.~\ref{transition}(b). In this step, the states $|ee\rangle$ and $|gg\rangle$ constitute the assistant subspace and the states $|eg\rangle$ and $|ge\rangle$ span the working subspace.

Similar to Eqs.~(\ref{AncibaseGeneral_high})-(\ref{AncibaseGeneral_across}) for $M=2$ and $N=2$, the ancillary bases in this case are found to be
\begin{equation}\label{AncillaryGHZ}
\begin{aligned}
|\tilde{\mu}_0(t)\rangle&=\sin\tilde{\theta}_0(t)|ee\rangle+\cos\tilde{\theta}_0(t)e^{-i\tilde{\alpha}_0(t)}|gg\rangle,\\
|\mu_0(t)\rangle&=\cos\theta_0(t)|eg\rangle-\sin\theta_0(t)e^{-i\alpha_0(t)}|ge\rangle,\\
|\mu_1(t)\rangle&=\cos\phi(t)|b_0(t)\rangle-\sin\phi(t)e^{-i\alpha(t)}|\tilde{b}_0(t)\rangle,\\
|\mu_2(t)\rangle&=\sin\phi(t)|b_0(t)\rangle+\cos\phi(t)e^{-i\alpha(t)}|\tilde{b}_0(t)\rangle,
\end{aligned}
\end{equation}
where $\tilde{\alpha}_0(t)$, $\alpha_0(t)$, $\alpha(t)$, $\tilde{\theta}_0(t)$, $\theta_0(t)$, and $\phi(t)$ are the time-dependent parameters. Due to Eq.~(\ref{brightGeneral}), the bright states are
\begin{equation}\label{brightGHZtwo}
\begin{aligned}
|\tilde{b}_0(t)\rangle&\equiv\cos\tilde{\theta}_0(t)|ee\rangle-\sin\tilde{\theta}_0(t)e^{-i\tilde{\alpha}_0(t)}|gg\rangle,\\
|b_0(t)\rangle&\equiv\sin\theta_0(t)|eg\rangle+\cos\theta_0(t)e^{-i\alpha_0(t)}|ge\rangle.
\end{aligned}
\end{equation}

Substituting the ancillary bases $|\mu_1(t)\rangle$ and $|\mu_2(t)\rangle$ of Eq.~(\ref{AncillaryGHZ}) into Eq.~(\ref{von}) with the effective Hamiltonian~(\ref{HameffGHZ}), we have
\begin{equation}\label{conditionGHZ_phase}
\begin{aligned}
&\varphi_0(t)=\varphi(t)+\tilde{\alpha}_0(t)+\alpha_0(t),\\
&\varphi_1(t)=\varphi(t),\\
&\Omega_0(t)=\Omega(t)\cos\tilde{\theta}_0(t)\cos\theta_0(t),\\
&\Omega_1(t)=\Omega(t)\cos\tilde{\theta}_0(t)\sin\theta_0(t),\\
&\dot{\alpha}(t)=-2\dot{\phi}(t)\cot\left[\varphi(t)+\alpha(t)\right]\cot2\phi(t),
\end{aligned}
\end{equation}
and we find $\varphi(t)=\pi/2$, $\tilde{\theta}_0(t)=\pi/4$ and $\theta_0(t)=\pi/4$. Under the conditions in Eq.~(\ref{conditionGHZ_phase}), $|\tilde{\mu}_0(t)\rangle$ and $|\mu_0(t)\rangle$ in Eq.~(\ref{AncillaryGHZ}) are dark states, while $|\mu_1(t)\rangle$ and $|\mu_2(t)\rangle$ can be used as the nonadiabatic passages. We assume that Step $2$ lasts the same period $T$ as Step $1$. When $\alpha_0(t)=0$, $\tilde{\alpha}_0(t)=0$, $\alpha(t)=\pi$, $\phi(T)=0$, and $\phi(2T)=\pi/2$, the single-excitation Bell state $(|eg\rangle+|ge\rangle)/\sqrt{2}$ is found to be converted to the double-excitation Bell state $(|ee\rangle-|gg\rangle)/\sqrt{2}$ via the path $|\mu_1(t)\rangle$. In addition, by using proper boundary conditions of $\alpha_0(t)$, $\tilde{\alpha}_0(t)$, $\alpha(t)$ and $\phi(t)$, the state transfer in Step $2$ can be exactly inverted. For example, when $\alpha_0(t)=0$, $\tilde{\alpha}_0(t)=0$, $\alpha(t)=\pi$, $\phi(T)=\pi/2$, and $\phi(2T)=0$, if the system is prepared as $(|ee\rangle-|gg\rangle)/\sqrt{2}$ when $t=T$, it can be converted to $(|eg\rangle+|ge\rangle)/\sqrt{2}$ when $t=2T$ via the same path $|\mu_1(t)\rangle$.

\begin{figure}[htbp]
\centering
\includegraphics[width=0.9\linewidth]{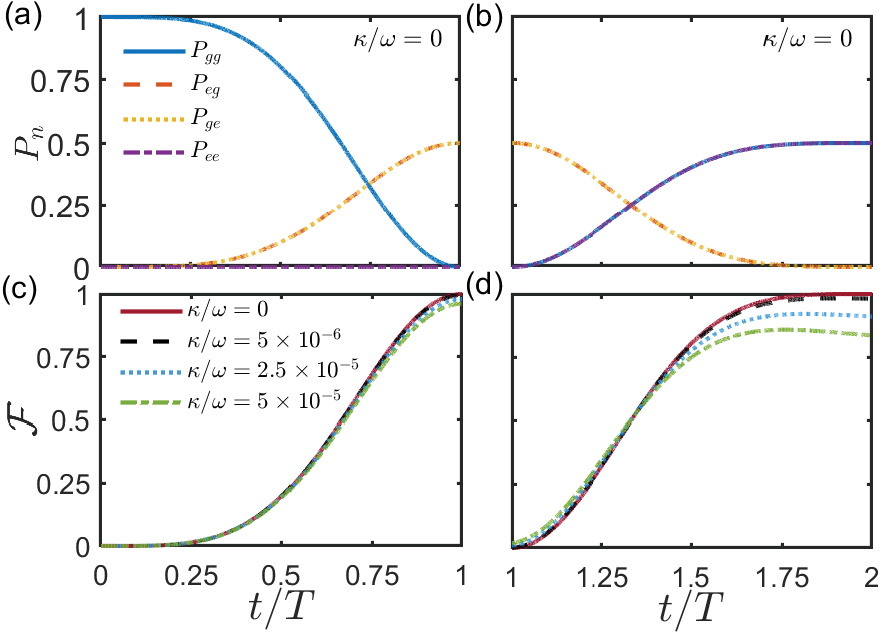}
\caption{(a) and (b) Population dynamics for closed system $P_n$, $n=gg, eg, ge, ee$, during Step $1$ and Step $2$, respectively. (c) and (d) Target-state fidelity dynamics $\mathcal{F}$ about the single-excitation Bell state $(|eg\rangle+|ge\rangle)/\sqrt{2}$ and the double-excitation Bell state $(|ee\rangle-|gg\rangle)/\sqrt{2}$, respectively, for both closed and open systems. In (a) and (c), $\omega_0=-\omega+J$, $J/\omega=0.1$, $\theta_0(t)=\pi/4$, $\varphi(t)=\pi/2$, $\alpha_0(t)=0$, $\alpha(t)=0$, $\phi(t)=(\pi/2)\cos[\pi t/(2T)]$, and the Rabi frequencies $\Omega_n(t)$ and the phases $\varphi_n(t)$ are set according to Eq.~(\ref{conditionBell}). In (b) and (d), $\omega_0=-\omega$ and $J/\omega=0$, $\varphi(t)=\pi/2$, $\alpha_0(t)=0$, $\tilde{\alpha}_0(t)=0$, $\alpha(t)=\pi$, $\phi(t)=(\pi/2)\cos[\pi t/(2T)]$, and $\Omega_n(t)$ and $\varphi_n(t)$ are set according to Eq.~(\ref{conditionGHZ_phase}).}\label{Bell}
\end{figure}

The performance in generation of Bell states can be measured by the population $P_n(t)\equiv\langle n|\rho(t)|n\rangle$ over the product states $|n\rangle$, $n=gg, eg, ge, ee$, and the target-state fidelity $\mathcal{F}(t)\equiv\langle\psi_{\rm Bell}|\rho(t)|\psi_{\rm Bell}\rangle$, where $\rho(t)$ is the time-evolved density matrix for the system qubits. In Step $1$, we have $|\psi_{\rm Bell}\rangle=(|eg\rangle+|ge\rangle)/\sqrt{2}$; and in Step $2$, we have $|\psi_{\rm Bell}\rangle=(|ee\rangle-|gg\rangle)/\sqrt{2}$. Initially the system state is $\rho(0)=|\mu_1(0)\rangle\langle\mu_1(0)|$. In the presence of external dissipation, the dynamics of population and fidelity can be estimated by the master equation~\cite{Carmichael1999statistical},
\begin{equation}\label{master}
\begin{aligned}
\frac{\partial \rho}{\partial t}&=-i[H_{\rm tot}(t), \rho]+\frac{\kappa}{2}\mathcal{L}(\sigma^-_1)+\frac{\kappa}{2}\mathcal{L}(\sigma^-_2).
\end{aligned}
\end{equation}
Here $H_{\rm tot}(t)$ is the full Hamiltonian in Eq.~(\ref{Hamfull}). $\mathcal{L}(o)$ is the Lindblad superoperator defined as $\mathcal{L}(o)\equiv2o\rho o^\dagger-o^\dagger o\rho-\rho o^\dagger o$~\cite{Scully1997quantum}, where $o=\sigma^-_1$ and $\sigma^-_2$. $\sigma^-_n\equiv|g\rangle_n\langle e|$ describes the dissipation channel of the $n$th qubit. The lifetime of the ASQ is typically in the order of $\sim20\mu$s~\cite{Hays2021Coherent,Pita2023Direct,Tosi2019Spin,Hays2020Continuous,
Wesdorp2023Dynamical,Wesdorp2024Microwave,Bargerbos2023Spectroscopy}. Then the decay rate is $\kappa/2\pi\sim10$ kHz. The running time for each step is set in the order of $T\sim1\mu$s in our numerical simulation.

The closed-system dynamics of population are demonstrated in Figs.~\ref{Bell}(a) and \ref{Bell}(b). The effect from external dissipation appears in the fidelity dynamics in Figs.~\ref{Bell}(c) and \ref{Bell}(d). During the first step $t\in[0,T]$ as shown in Fig.~\ref{Bell}(a), the population on $|gg\rangle$ is transferred equally to $|eg\rangle$ and $|ge\rangle$ and the state $|ee\rangle$ is never populated as described by $|\mu_1(t)\rangle$ in Eq.~(\ref{AncillaryBell}). Together with the solid line in Fig.~\ref{Bell}(c), by which a target state $|\psi_{\rm Bell}\rangle=(|eg\rangle+|ge\rangle)/\sqrt{2}$ is achieved in the end of Step $1$, it confirms that the system plotted in Fig.~\ref{transition}(a) can be faithfully simplified to be a $1+2$-dimensional system. In the presence of a dissipative noise, the fidelity at $t=T$ can be maintained as high as $\mathcal{F}=0.997$ under the experimentally practical decay rate $\kappa/\omega=5\times10^{-6}$~\cite{Pita2024Strong}, by which our protocol prevails over the existing one in the superconducting waveguide QED system~\cite{Zhang2023Generating} with the same target state and the same order of magnitude of noise. It is found in Fig.~\ref{Bell}(c) that $\mathcal{F}(T)=0.980$ for $\kappa/\omega=2.5\times10^{-5}$ and $\mathcal{F}(T)=0.962$ for $\kappa/\omega=5\times10^{-5}$. Our protocol can then tolerate larger decay rates.

During the second step, e.g., $t\in[T,2T]$, as shown in Fig.~\ref{Bell}(b), the states $|ee\rangle$ and $|gg\rangle$ are synchronously populated as the populations on the states $|eg\rangle$ and $|ge\rangle$ decrease with time. The relevant dynamics of target-state fidelity for the double-excitation Bell state $(|ee\rangle-|gg\rangle)/\sqrt{2}$ is shown in Fig.~\ref{Bell}(d), which seems more fragile to the dissipative noise than that for the single-excitation Bell state. When $t=2T$, the fidelity can remain nearly unit for $\kappa/\omega=5\times10^{-6}$. And it reduces to $\mathcal{F}=0.912$ for $\kappa/\omega=2.5\times10^{-5}$ and $\mathcal{F}=0.836$ for $\kappa/J=5\times10^{-5}$.

\begin{figure}[htbp]
\centering
\includegraphics[width=0.9\linewidth]{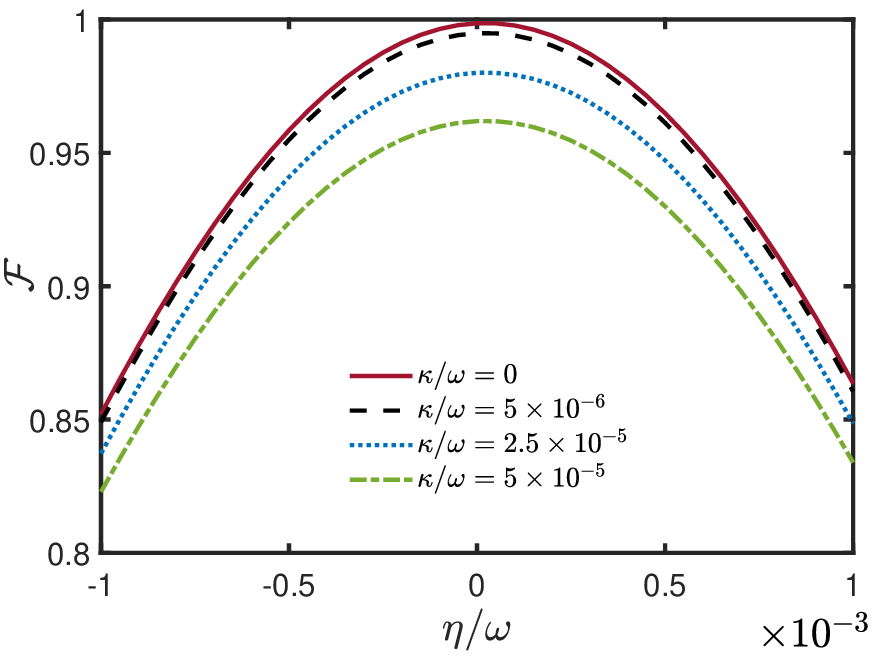}
\caption{Target-state fidelity $\mathcal{F}$ about the single-excitation Bell state $(|eg\rangle+|ge\rangle)/\sqrt{2}$ vs the deviation $\eta$ in the driving frequency for both closed and open systems. The other parameters are set the same as Figs.~\ref{Bell}(a) and \ref{Bell}(c).}\label{Error}
\end{figure}

In the presence of the deviation in the driving frequency, i.e.,  $\omega_0\rightarrow\omega_0+\eta$ with $\eta$ the error magnitude, the rotated Hamiltonian in Eq.~(\ref{HamqubRot}) becomes
\begin{equation}\label{HamrotDev}
\begin{aligned}
&H_{\rm rot}'(t)=\\
&e^{i(\omega+\omega_0+J)t}e^{i\eta t}\times[\Omega_0(t)e^{i\varphi_0(t)}(|ee\rangle\langle ge|+e^{-i2Jt}|eg\rangle\langle gg|)\\
&+\Omega_1(t)e^{i\varphi_1(t)}(|ee\rangle\langle eg|+e^{-i2Jt}|ge\rangle\langle gg|)]+{\rm H.c.}.
\end{aligned}
\end{equation}
The effect of this systematic error on the fidelity of the single-excitation Bell state is shown in Fig.~\ref{Error}, using the master equation~(\ref{master}) with the perturbated Hamiltonian~(\ref{HamrotDev}). It is found that the fidelity for the open system with $\kappa/\omega=5\times10^{-6}$ is still the same as the closed system. In particular, for the closed system, i.e., $\kappa/\omega=0$, the fidelity is $\mathcal{F}(T)=0.958$ when $\eta/\omega=-0.5\times10^{-3}$ and $\mathcal{F}(T)=0.964$ when $\eta/\omega=0.5\times10^{-3}$. In the case of $\kappa/\omega=5\times10^{-6}$, $\mathcal{F}(T)=0.955$ when $\eta/\omega=-0.5\times10^{-3}$ and $\mathcal{F}(T)=0.961$ when $\eta/\omega=0.5\times10^{-3}$. The largest distinction occurs when $\eta/\omega=0$. For $\kappa/\omega=0$, $\mathcal{F}$ is unit; and for $\kappa/\omega=5\times10^{-6}$, $\mathcal{F}=0.995$. It means that the current application is insensitive to certain systematical error.

\subsection{Three-qubit system}\label{Threeentangle}

With an extended Hamiltonian as to that in Eq.~(\ref{Hamfull}), a GHZ state of three qubits can be generated via our universal control theory with three steps. In particular, we consider that the first and third qubits can be longitudinally coupled to the second one with the same coupling strength $J$ and each qubit is driven by a local laser field. The full Hamiltonian reads
\begin{equation}\label{Hamthree}
H(t)=H+H_d(t)
\end{equation}
with
\begin{equation}\label{H0HIHd}
\begin{aligned}
H&=\frac{\omega}{2}(\sigma_1^z+\sigma_2^z+\sigma_3^z)+\frac{J}{2}(\sigma_1^z\sigma_2^z+\sigma_2^z\sigma_3^z),\\
H_d(t)&=e^{i\omega_0t}[\Omega_0(t)e^{i\varphi_0(t)}|e\rangle_1\langle g|\\
+&\Omega_1(t)e^{i\varphi_1(t)}|e\rangle_2\langle g|+\Omega_2(t)e^{i\varphi_2(t)}|e\rangle_3\langle g|]+{\rm H.c.},
\end{aligned}
\end{equation}
where $\Omega_n(t)$ is the Rabi frequency, $\omega_0$ is the common driving frequency, and $\varphi_n(t)$ is the driving phase. In the rotating frame with respect to $H$~(\ref{H0HIHd}), the full Hamiltonian~(\ref{Hamthree}) is rewritten as
\begin{equation}\label{HamThRot}
\begin{aligned}
&H_{\rm rot}(t)\\
&=e^{i(\omega+\omega_0-J)t}[\Omega_0(t)e^{i\varphi_0(t)}(e^{i2Jt}|ee\rangle_{12}\langle ge|\\
&+|eg\rangle_{12}\langle gg|)\otimes\mathcal{I}_3+\Omega_1(t)e^{i(Jt+\varphi_1(t))}(e^{i2Jt}|eee\rangle\langle ege|\\
&+|eeg\rangle\langle egg|+|gee\rangle\langle gge|+e^{-i2Jt}|geg\rangle\langle ggg|)\\
&+\Omega_2(t)e^{i\varphi_2(t)}\mathcal{I}_1\otimes(e^{i2Jt}|ee\rangle_{23}\langle eg|+|ge\rangle_{23}\langle gg|)]+{\rm H.c.},
\end{aligned}
\end{equation}
where $\mathcal{I}_n$ denotes the identity operator of the $n$th qubit.

\begin{figure}[htbp]
\centering
\includegraphics[width=0.9\linewidth]{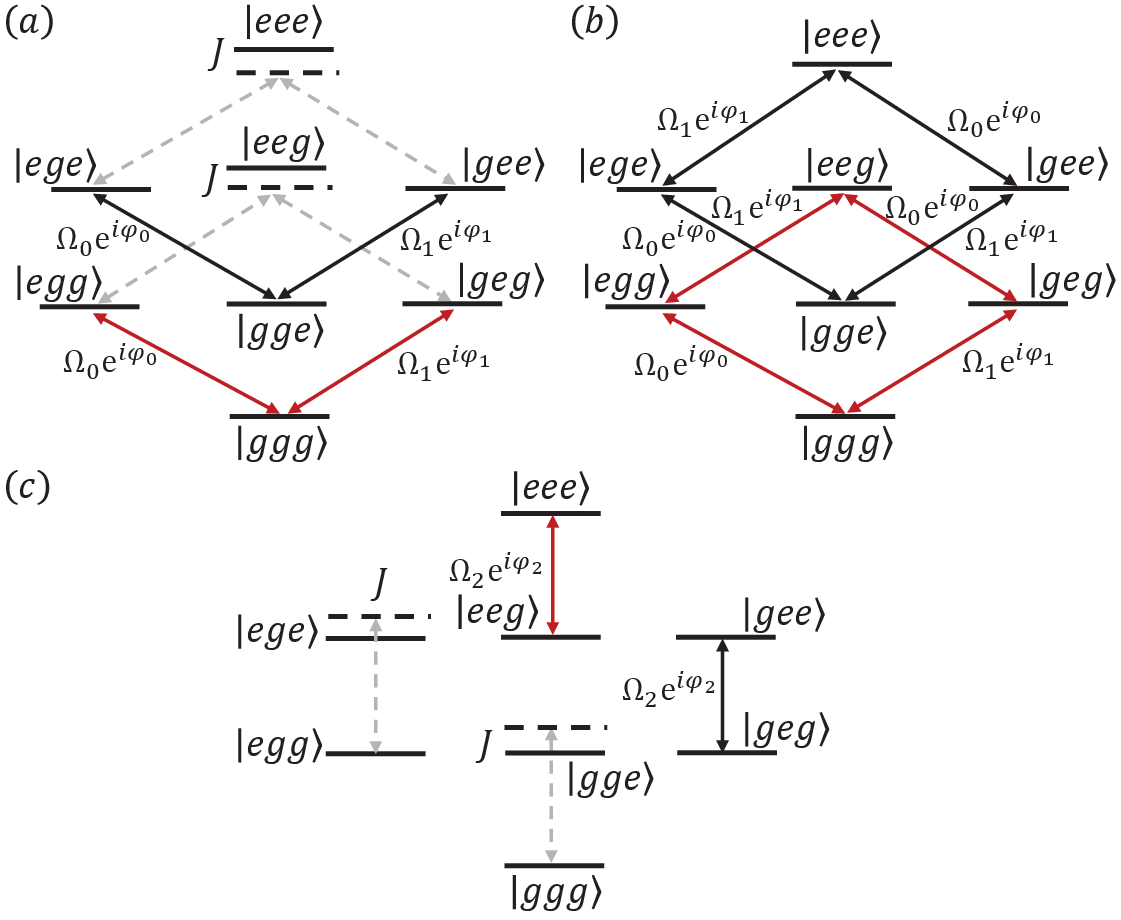}
\caption{Step-wise transition diagrams for the longitudinally coupled three-qubit system for (a) the interaction between the second qubit and the third qubit is turned off, the Rabi frequency $\Omega_2(t)=0$, the driving frequency $\omega_0=-\omega+J$, and $J\gg\Omega_0(t),\Omega_1(t)$; (b) when $\Omega_2(t)=0$, $\omega_0=-\omega$, and $J=0$; and (c) when the interaction between the first qubit and the second qubit is turned off, $\Omega_0(t)=0$, $\Omega_1(t)=0$, $\omega_0=-\omega-J$, and $J\gg\Omega_2(t)$. The transitions described by the gray dashed lines are suppressed due to the strong longitudinal interaction, and the transitions about the generation of the GHZ state are distinguished by the red solid lines.}\label{transitionthree}
\end{figure}

\emph{Step 1}--- Similar to the first step in Sec.~\ref{Twoentangle}, this step contributes to pushing the system from the ground state $|ggg\rangle$ to the first intermediate state $(|eg\rangle+|ge\rangle)/\sqrt{2}\otimes|g\rangle$, during which the driving field on the third qubit and the coupling between the second and third qubits are switched off, i.e., $\Omega_2(t)=0$. Under the driving frequency $\omega_0=-\omega+J$ and the strong coupling strength $J\gg\Omega_0(t), \Omega_1(t)$, the Hamiltonian $H_{\rm rot}(t)$~(\ref{HamThRot}) can be effectively written as
\begin{equation}\label{HamThEff1}
\begin{aligned}
H_{\rm eff}^{(1)}&=\Omega_0(t)e^{i\varphi_0}(|ege\rangle\langle gge|+|egg\rangle\langle ggg|)\\
&+\Omega_1(t)e^{i\varphi_1}(|gee\rangle\langle gge|+|geg\rangle\langle ggg|)+{\rm H.c.}
\end{aligned}
\end{equation}
under RWA. In Fig.~\ref{transitionthree}(a), the transitions $|eee\rangle\leftrightarrow|ege\rangle$, $|eee\rangle\leftrightarrow|gee\rangle$, $|eeg\rangle\leftrightarrow|egg\rangle$, and $|eeg\rangle\leftrightarrow|geg\rangle$ indicated by the gray dashed lines are suppressed due to the strong longitudinal interaction. The states $|ggg\rangle$, $|egg\rangle$, and $|geg\rangle$ then constitute a $1+2$-dimensional subspace, which has been illustrated in Secs.~\ref{illustrative} or ~\ref{Twoentangle}. Similar to Eqs.~(\ref{AncillaryBell_work}) and (\ref{AncillaryBell}) for the two-qubit system, the ancillary bases in this step can be written as
\begin{equation}\label{AncillaryBellTh}
\begin{aligned}
|\mu_0(t)\rangle&=\cos\theta_0(t)|egg\rangle-\sin\theta_0(t)e^{-i\alpha_0(t)}|geg\rangle,\\
|\mu_1(t)\rangle&=\cos\phi(t)|b_0(t)\rangle-\sin\phi(t)e^{-i\alpha(t)}|ggg\rangle,\\
|\mu_2(t)\rangle&=\sin\phi(t)|b_0(t)\rangle+\cos\phi(t)e^{-i\alpha(t)}|ggg\rangle,
\end{aligned}
\end{equation}
where $|b_0(t)\rangle=\sin\theta_0(t)|egg\rangle+\cos\theta_0(t)e^{-i\alpha_0(t)}|geg\rangle$. Under the same conditions as in Eq.~(\ref{conditionBell}), $|\mu_1(t)\rangle$ and $|\mu_2(t)\rangle$ can be used as the nonadiabatic passages. With no loss of generality, the system can evolve as $|ggg\rangle\rightarrow\sin\theta_0(t)|egg\rangle+\cos\theta_0(t)e^{-i\alpha_0(t)}|geg\rangle$ via $|\mu_1(t)\rangle$ under the boundary conditions $\alpha(t)=\pi$, $\phi(0)=\pi/2$, and $\phi(T)=0$. If we further set $\alpha_0(t)=0$ and $\theta_0(t)=\pi/4$, then the system can evolve to $(|egg\rangle+|geg\rangle)/\sqrt{2}$.

\emph{Step 2}--- Again similar to the second step in Sec.~\ref{Twoentangle}, we convert the system from  $(|egg\rangle+|geg\rangle)/\sqrt{2}$ to $(|eeg\rangle-|ggg\rangle)/\sqrt{2}$ during this step by setting $J=0$. Under the driving frequency $\omega_0=-\omega$ on the first two qubits, the Hamiltonian~(\ref{HamThRot}) can be written as
\begin{equation}\label{HamThEff2}
\begin{aligned}
H_{\rm eff}^{(2)}&=\Omega_0(t)e^{i\varphi_0}(|eee\rangle\langle gee|+|eeg\rangle\langle geg|\\
&+|ege\rangle\langle gge|+|egg\rangle\langle ggg|)\\
&+\Omega_1(t)e^{i\varphi_1}(|eee\rangle\langle ege|+|eeg\rangle\langle egg|\\
&+|gee\rangle\langle gge|+|geg\rangle\langle ggg|)+{\rm H.c.}.
\end{aligned}
\end{equation}
In Fig.~\ref{transitionthree}(b), we therefore have double $2+2$-dimensional subspaces. The first one is $\{|ggg\rangle, |eeg\rangle, |egg\rangle, |geg\rangle\}$ and the second one is $\{|gge\rangle, |eee\rangle, |ege\rangle, |gee\rangle\}$. Since in the end of the last step, the system is only populated on $|egg\rangle$ and $|geg\rangle$, we here focus on the first subspace which can be further divided into the assistant subspace spanned by $|eeg\rangle$ and $|ggg\rangle$ and the working subspace spanned by $|egg\rangle$ and $|geg\rangle$. Similar to Eq.~(\ref{AncillaryGHZ}), the ancillary bases can be chosen as
\begin{equation}\label{AncillaryGHZTh}
\begin{aligned}
|\tilde{\mu}_0(t)\rangle&=\sin\tilde{\theta}_0(t)|eeg\rangle+\cos\tilde{\theta}_0(t)e^{-i\tilde{\alpha}_0(t)}|ggg\rangle,\\
|\mu_0(t)\rangle&=\cos\theta_0(t)|egg\rangle-\sin\theta_0(t)e^{-i\alpha_0(t)}|geg\rangle,\\
|\mu_1(t)\rangle&=\cos\phi(t)|b_0(t)\rangle-\sin\phi(t)e^{-i\alpha(t)}|\tilde{b}_0(t)\rangle, \\
|\mu_2(t)\rangle&=\sin\phi(t)|b_0(t)\rangle+\cos\phi(t)e^{-i\alpha(t)}|\tilde{b}_0(t)\rangle,
\end{aligned}
\end{equation}
where the bright states are
\begin{equation}\label{brightGHzthree}
\begin{aligned}
|\tilde{b}_0(t)\rangle&=\cos\tilde{\theta}_0(t)|eeg\rangle-\sin\tilde{\theta}_0(t)e^{-i\tilde{\alpha}_0(t)}|ggg\rangle,\\
|b_0(t)\rangle&=\sin\theta_0(t)|egg\rangle+\cos\theta_0(t)e^{-i\alpha_0(t)}|geg\rangle.
\end{aligned}
\end{equation}
Under the conditions in Eq.~(\ref{conditionGHZ_phase}) with $\alpha_0(t)=0$, $\tilde{\alpha}_0(t)=0$, $\alpha(t)=\pi$, $\phi(T)=0$, and $\phi(2T)=\pi/2$, the states can be converted from the state $(|egg\rangle+|geg\rangle)/\sqrt{2}$ to the state $(|eeg\rangle-|ggg\rangle)/\sqrt{2}$ via $|\mu_1(t)\rangle$.

\emph{Step 3}--- In this step, we can evolve the system from the intermediated state $(|eeg\rangle-|ggg\rangle)/\sqrt{2}$ to the GHZ state $(|eee\rangle-|ggg\rangle)/\sqrt{2}$ by turning off the coupling between the first and second qubits and the driving fields on them, i.e., $\Omega_0(t)=0$ and $\Omega_1(t)=0$. Turning on the driving on the third qubit with the driving frequency $\omega_0=-\omega-J$ with $J\gg\Omega_2(t)$ between the second and third qubits, the Hamiltonian in Eq.~(\ref{HamThRot}) reduces to
\begin{equation}\label{HamThEff3}
H_{\rm eff}^{(3)}=\Omega_2(t)e^{i\varphi_2(t)}(|eee\rangle\langle eeg|+|gee\rangle\langle geg|)+{\rm H.c.}
\end{equation}
under RWA. The associated transition diagram is shown in Fig.~\ref{transitionthree}(c). It is found that transition $|ggg\rangle\leftrightarrow|gge\rangle$ is averaged out in the system dynamics due to the strong longitudinal interaction. A previous protocol about the generation of a three-qubit GHZ state in a superconducting circuit system~\cite{Wei2006Generation} is based on the ultrastrong longitudinal interaction between the qubits. It demands $J\sim\omega$ and then is challenged in experiment. In contrast, we take advantage of a tunable driving frequency $\omega_0$ to avoid the rigorous requirement about the coupling strength.

We here focus on the subspace spanned by $|eeg\rangle$ and $|eee\rangle$. The two ancillary bases can be chosen as
\begin{equation}\label{AncillaryTh3}
\begin{aligned}
|\mu_0(t)\rangle&=\cos\phi(t)|eeg\rangle-\sin\phi(t)e^{-i\alpha(t)}|eee\rangle,\\
|\mu_1(t)\rangle&=\sin\phi(t)|eeg\rangle+\cos\phi(t)e^{-i\alpha(t)}|eee\rangle.
\end{aligned}
\end{equation}
Substituting them into the von Neumann equation~(\ref{von}) with the Hamiltonian in Eq.~(\ref{HamThEff3}), we have the phase and the Rabi frequency as
\begin{equation}\label{conditionTh3}
\begin{aligned}
\Omega_2(t)&=-\frac{\dot{\phi}(t)}{\sin\left[\varphi_2(t)+\alpha(t)\right]},\\
\dot{\alpha}(t)&=-2\dot{\phi}(t)\cot\left[\varphi_2(t)+\alpha(t)\right]\cot2\phi(t).
\end{aligned}
\end{equation}
Under the conditions in Eq.~(\ref{conditionTh3}) with $\alpha(t)=\pi$, $\phi(2T)=0$, and $\phi(3T)=\pi/2$, the system can be transferred to the GHZ state $(|eee\rangle-|ggg\rangle)/\sqrt{2}$ via $|\mu_0(t)\rangle$.

\begin{figure}[htbp]
\centering
\includegraphics[width=0.8\linewidth]{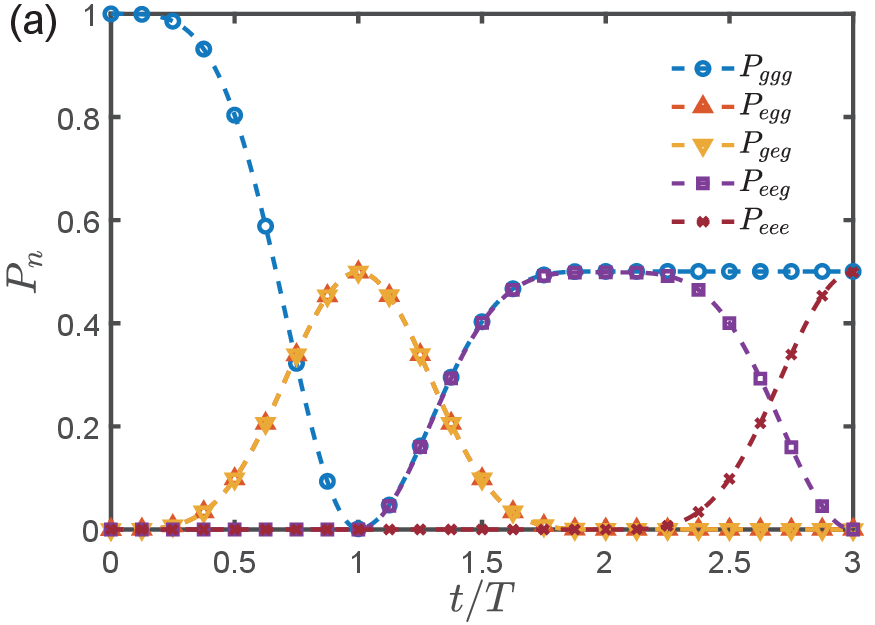}
\includegraphics[width=0.8\linewidth]{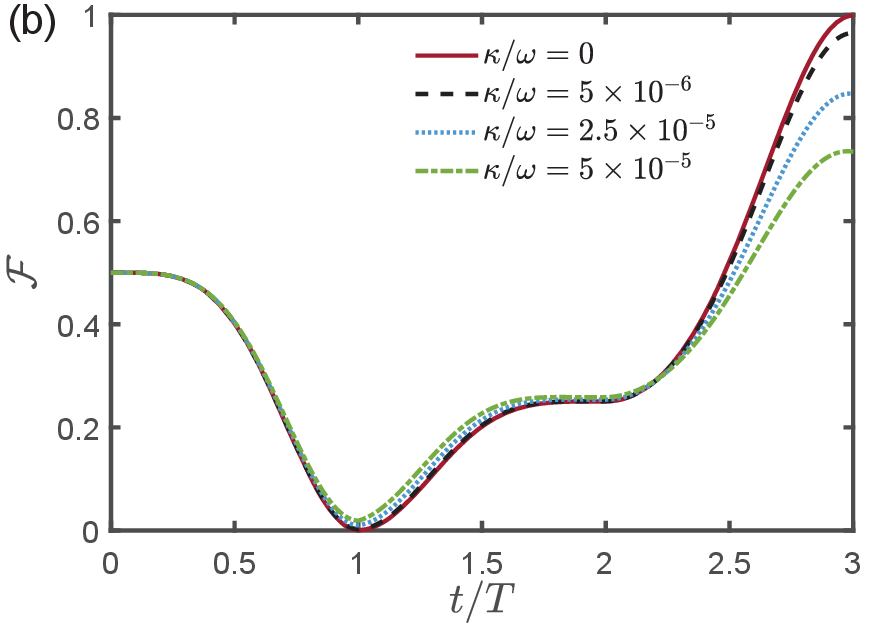}
\caption{(a) Populations dynamics for the closed system, and (b) target-state fidelity dynamics about the three-qubit GHZ state. When $t\in[0,2T]$, the parameters are set the same as those in Fig.~\ref{Bell}. When $t\in[2T,3T]$, $\alpha(t)=\pi$, $\varphi(t)=\pi/2$, $\phi(t)=-(\pi/2)\cos[\pi t/(2T)]$, and the Rabi frequency $\Omega_2(t)$ and the phase $\varphi_2(t)$ are set as Eq.~(\ref{conditionTh3}).}\label{GHZ}
\end{figure}

Similar to Eq.~(\ref{master}), the three-qubit system dynamics $\rho(t)$ in the presence of the dissipation noise can be calculated by the master equation
\begin{equation}\label{masterthree}
\frac{\partial \rho}{\partial t}=-i[H(t),\rho]+\frac{\kappa}{2}\mathcal{L}(\sigma^-_1)+\frac{\kappa}{2}\mathcal{L}(\sigma^-_2)
+\frac{\kappa}{2}\mathcal{L}(\sigma^-_3),
\end{equation}
where $H(t)$ is the full Hamiltonian~(\ref{Hamthree}) and the dissipation rates for the three qubits are set as the same value $\kappa$ for simplicity.

The dynamics about the level populations $P_n$ ($n=ggg,egg,geg,eeg,eee$ for all the involved levels in the above three steps) and the target-state fidelity $\mathcal{F}=\langle\psi_{\rm GHZ}|\rho(t)|\psi_{\rm GHZ}\rangle$ with $|\psi_{\rm GHZ}\rangle=(|eee\rangle-|ggg\rangle)/\sqrt{2}$ are demonstrated in Figs.~\ref{GHZ}(a) and \ref{GHZ}(b), respectively. During the first step $t\in[0,T]$ as shown in Fig.~\ref{GHZ}(a), the population on the ground state $|ggg\rangle$ is equally transferred to the levels $|egg\rangle$ and $|geg\rangle$. The other states are not populated during this process. During the second step $t\in[T,2T]$, the populations on the states $|egg\rangle$ and $|geg\rangle$ can be synchronically converted to the states $|ggg\rangle$ and $|eeg\rangle$ until they are equally populated as $P_{ggg}=P_{eeg}=0.5$. Then, during the third step $t\in[2T,3T]$, the population on the state $|eeg\rangle$ can be fully transferred to the state $|eee\rangle$ and that on the state $|ggg\rangle$ remains invariant due to the fact that $|ggg\rangle$ is decoupled from the system dynamics indicated by Fig.~\ref{transitionthree}(c). Figure~\ref{GHZ}(b) shows the decoherence effect on the state fidelity. In the end of the control with $t=3T$, one can have a faithful GHZ state $(|eee\rangle-|ggg\rangle)/\sqrt{2}$ in the absence of dissipation $\kappa/\omega=0$. The fidelity decreases with increasing $\kappa$. It is over $\mathcal{F}=0.965$ when $\kappa/\omega=5\times10^{-6}$ and declines to $\mathcal{F}=0.735$ when $\kappa/\omega=5\times10^{-5}$.

\subsection{$N$-qubit system}

We can create the $N$-qubit GHZ state on a generalized Hamiltonian similar to that in Eq.~(\ref{Hamthree}) within $N$ steps. In particular, we require that the $n$th qubit, $1\le n\le N-1$, is longitudinally interacted with the $(n+1)$th one with the same coupling strength $J$, and each qubit can be driven by a local field of the same frequency. The full Hamiltonian can therefore be written as
\begin{equation}\label{HamNqubit}
H(t)=H_0+H_I+H_d(t),
\end{equation}
where
\begin{equation}\label{H0HIHdN}
\begin{aligned}
H_0&=\frac{\omega}{2}\sum_{n=1}^N\sigma_n^z,\\
H_I&=\frac{J}{2}\sum_{n=1}^{N-1}\sigma_n^z\sigma_{n+1}^z,\\
H_d(t)&=\sum_{n=0}^{N-1}\Omega_n(t)e^{i\left[\omega_0t+\varphi_n(t)\right]}|e\rangle_{n+1}\langle g|+{\rm H.c.},
\end{aligned}
\end{equation}
where $\Omega_n(t)$ is the Rabi frequency, $\omega_0(t)$ is the common driving frequency of $N$ laser fields, and $\varphi_n(t)$ is the time-dependent phase. In the rotating frame with respect to $H_0$, the full Hamiltonian is
\begin{equation}\label{HamintN}
H_{\rm int}(t)=H_I+\sum_{n=0}^{N-1}\Omega_n(t)e^{i\left[(\omega+\omega_0)t+\varphi_n\right]}|e\rangle_{n+1}\langle g|+{\rm H.c.}.
\end{equation}

\emph{Step 1}--- Similar to the first step in Sec.~\ref{Twoentangle} or Sec.~\ref{Threeentangle}, this step targets to prepare the system to be the intermediate state $(|eg\rangle+|ge\rangle)/\sqrt{2}\otimes|g\rangle^{\otimes(N-2)}$ from the ground state $|g\rangle^{\otimes N}$. We turn on merely the driving fields on the first two qubits and their coupling and set $\omega_0=-\omega+J$ with $J\gg\{\Omega_0, \Omega_1\}$. The rotating Hamiltonian with respect to $H_I$ can then be expressed as
\begin{equation}\label{HameffN1}
\begin{aligned}
H_{\rm eff}^{(1)}&=\Omega_0(t)e^{i\varphi_0(t)}|eg\cdots gg\rangle\langle gg\cdots gg|\\
&+\Omega_1(t)e^{i\varphi_1(t)}|ge\cdots gg\rangle\langle gg\cdots gg|+{\rm H.c.}
\end{aligned}
\end{equation}
under RWA. Here, the states $|gg\rangle\otimes|g\rangle^{\otimes(N-2)}$, $|ge\rangle\otimes|g\rangle^{\otimes(N-2)}$, and $|eg\rangle\otimes|g\rangle^{\otimes(N-2)}$ constitute a $1+2$-dimensional subspace. The ancillary bases in this step can be obtained by using Eq.~(\ref{AncillaryBellTh}) for the three-qubit system:
\begin{equation}\label{AncillaryBellN}
\begin{aligned}
|\mu_0(t)\rangle&=\cos\theta_0(t)|eg\cdots gg\rangle-\sin\theta_0(t)e^{-i\alpha_0(t)}|ge\cdots gg\rangle,\\
|\mu_1(t)\rangle&=\cos\phi(t)|b_0(t)\rangle-\sin\phi(t)e^{-i\alpha(t)}|gg\cdots gg\rangle,\\
|\mu_2(t)\rangle&=\sin\phi(t)|b_0(t)\rangle+\cos\phi(t)e^{-i\alpha(t)}|gg\cdots gg\rangle,
\end{aligned}
\end{equation}
where the bright state is $|b_0(t)\rangle=\sin\theta_0(t)|eg\cdots gg\rangle+\cos\theta_0(t)e^{-i\alpha_0(t)}|ge\cdots gg\rangle$. When the driving fields follow Eq.~(\ref{conditionBell}) with the boundary conditions $\alpha_0(t)=0$, $\alpha(t)=\pi$, $\phi(0)=\pi/2$, and $\phi(T)=0$, the system can evolve from $|gg\rangle\otimes|g\rangle^{\otimes(N-2)}$ to $(|eg\rangle+|ge\rangle)/\sqrt{2}\otimes|g\rangle^{\otimes(N-2)}$ via the passage $|\mu_1(t)\rangle$.

\emph{Step 2}--- Again similar to the second step in Secs.~\ref{Twoentangle} or \ref{Threeentangle}, this step converts the system from $(|eg\rangle+|ge\rangle)/\sqrt{2}\otimes|g\rangle^{\otimes(N-2)}$ to  $(|ee\rangle-|gg\rangle)/\sqrt{2}\otimes|g\rangle^{\otimes(N-2)}$. We turn off the longitudinal interaction among the first two qubits and set the driving frequency as $\omega_0=-\omega$. The effective Hamiltonian then becomes
\begin{equation}\label{HameffN2}
\begin{aligned}
H_{\rm eff}^{(2)}&=\Omega_0(t)e^{i\varphi_0(t)}(|eg\cdots gg\rangle\langle gg\cdots gg|\\
&+|ee\cdots gg\rangle\langle ge\cdots gg|)\\
&+\Omega_1(t)e^{i\varphi_1(t)}(|ge\cdots gg\rangle\langle gg\cdots gg|\\
&+|ee\cdots gg\rangle\langle eg\cdots gg|)+{\rm H.c.},
\end{aligned}
\end{equation}
which forms a $2+2$-dimensional system. The ancillary bases can be written as
\begin{equation}\label{AncillaryGHZN}
\begin{aligned}
|\tilde{\mu}_0(t)\rangle&=\sin\tilde{\theta}_0(t)|ee\cdots gg\rangle+\cos\tilde{\theta}_0(t)e^{-i\tilde{\alpha}_0(t)}|gg\cdots gg\rangle,\\
|\mu_0(t)\rangle&=\cos\theta_0(t)|eg\cdots gg\rangle-\sin\theta_0(t)e^{-i\alpha_0(t)}|ge\cdots gg\rangle,\\
|\mu_1(t)\rangle&=\cos\phi(t)|b_0(t)\rangle-\sin\phi(t)e^{-i\alpha(t)}|\tilde{b}_0(t)\rangle,\\
|\mu_2(t)\rangle&=\sin\phi(t)|b_0(t)\rangle+\cos\phi(t)e^{-i\alpha(t)}|\tilde{b}_0(t)\rangle,
\end{aligned}
\end{equation}
where the bright states are
\begin{equation}\label{brightGHZN}
\begin{aligned}
|\tilde{b}_0(t)\rangle&\equiv\cos\tilde{\theta}_0(t)|ee\cdots gg\rangle-\sin\tilde{\theta}_0(t)e^{-i\tilde{\alpha}_0(t)}|gg\cdots gg\rangle,\\
|b_0(t)\rangle&\equiv\sin\theta_0(t)|eg\cdots gg\rangle+\cos\theta_0(t)e^{-i\tilde{\alpha}_0(t)}|ge\cdots gg\rangle.
\end{aligned}
\end{equation}
With the phases and Rabi frequencies in Eq.~(\ref{conditionGHZ_phase}) and the boundary conditions $\alpha_0(t)=0$, $\tilde{\alpha}_0(t)=0$, $\alpha(t)=\pi$, $\phi(T)=0$, and $\phi(2T)=\pi/2$, the system can be transferred to $(|ee\rangle-|gg\rangle)/\sqrt{2}\otimes|g\rangle^{\otimes(N-2)}$ via the passage $|\mu_1(t)\rangle$.

\emph{Step $k$}, $3\le k\le N$, --- Similar to the third step in Sec.~\ref{Threeentangle}, the $k$th step is used to transform the system from the state $(|e\rangle^{\otimes(k-1)}-|g\rangle^{\otimes(k-1)})/\sqrt{2}\otimes|g\rangle^{\otimes(N-k+1)}$ to the state $(|e\rangle^{\otimes k}-|g\rangle^{\otimes k})/\sqrt{2}\otimes|g\rangle^{\otimes(N-k)}$. In the $k$th step, we hold exclusively the longitudinal interaction between the $(k-1)$th and $k$th qubits and the driving fields on the $k$th qubit. With $\omega_0=-\omega-J$ and $J\gg\Omega_{k-1}(t)$, the rotating Hamiltonian with respect to $H_I$ can then be approximated by
\begin{equation}\label{HameffNk}
\begin{aligned}
H_{\rm eff}^{(k)}&=\Omega_{k-1}(t)e^{i\varphi_{k-1}(t)}\\
&\times|ee\cdots e_k\cdots gg\rangle\langle ee\cdots g_k\cdots gg\rangle+{\rm H.c.},
\end{aligned}
\end{equation}
where $|e_k\rangle$ and $|g_k\rangle$ denote the excited state and the ground state for the $k$th qubit, respectively.

Similar to the $1+1$-dimensional system in Secs.~\ref{illustrative} and \ref{Threeentangle}, the ancillary bases for the nonadiabatic passage can be chosen from
\begin{equation}\label{AncillaryThN}
\begin{aligned}
|\mu_0(t)\rangle^{(k)}&=\cos\phi(t)|ee\cdots e_k\cdots gg\rangle\\
&-\sin\phi(t)e^{-i\alpha(t)}|ee\cdots g_k\cdots gg\rangle,\\
|\mu_1(t)\rangle^{(k)}&=\sin\phi(t)|ee\cdots e_k\cdots gg\rangle\\
&+\cos\phi(t)e^{-i\alpha(t)}|ee\cdots g_k\cdots gg\rangle,
\end{aligned}
\end{equation}
where the superscript $k$ indicates the $k$th step. With the driving fields under the conditions in Eq.~(\ref{conditionTh3}), the system can evolve from the state $(|e\rangle^{\otimes(k-1)}-|g\rangle^{\otimes(k-1)})/\sqrt{2}\otimes|g\rangle^{\otimes(N-k+1)}$ to the state $(|e\rangle^{\otimes k}-|g\rangle^{\otimes k})/\sqrt{2}\otimes|g\rangle^{\otimes(N-k)}$ via the passage $|\mu_0(t)\rangle^{(k)}$ by setting $\alpha(t)=\pi$, $\phi[(k-1)T]=\pi/2$, and $\phi(kT)=\pi$. Here $T$ is the running time for every step. With $N-2$ iterations of such a controlled evolution, the system will end up with an $N$-qubit GHZ state as $(|e\rangle^{\otimes N}-|g\rangle^{\otimes N})/\sqrt{2}$.

\begin{figure}[htbp]
\centering
\includegraphics[width=0.9\linewidth]{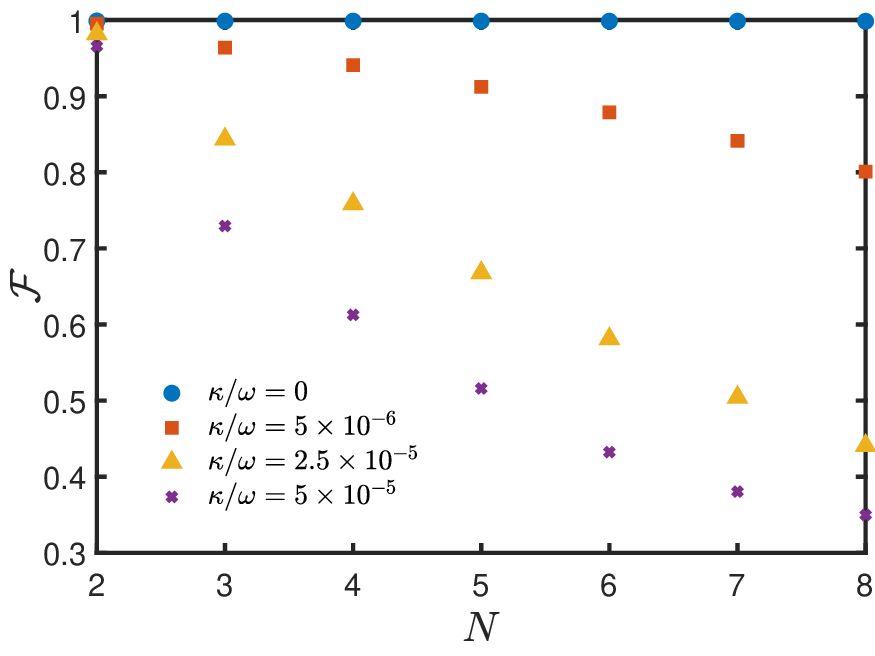}
\caption{Target-state fidelity $\mathcal{F}$ about the $N$-qubit GHZ state $[|e\rangle^{\otimes N}-(-1)^N|g\rangle^{\otimes N}]/\sqrt{2}$ vs the system size $N$, for both closed and open systems. The parameters for the first two steps are set the same as Fig.~\ref{Bell}(a) and (c); and the driving fields for the $k$th step, $3\le k\le N$, are under the condition in Eq.~(\ref{conditionTh3}) with $\alpha(t)=\pi$ and $\phi(t)=(\pi/2)\cos\left\{\pi\left[t-(k-3)T\right]/(2T)\right\}$.}\label{Deep}
\end{figure}

Similar to Eqs.~(\ref{master}) and (\ref{masterthree}), the $N$-qubit system dynamics $\rho(t)$ in the presence of the dissipative noise can be calculated by the master equation
\begin{equation}\label{masterN}
\frac{\partial \rho}{\partial t}=-i[H(t),\rho]+\frac{\kappa}{2}\sum_{n=1}^{N}\mathcal{L}(\sigma^-_n),
\end{equation}
where $H(t)$ is the full Hamiltonian~(\ref{HamNqubit}). In Fig.~\ref{Deep}, we demonstrate the target-state fidelity about the $N$-qubit GHZ state versus the qubit number $N$. It is found that our protocol is extendable for large-scale quantum applications. In the ideal situation, i.e., $\kappa/\omega=0$, the fidelity remains almost unit for all the GHZ states. When the dissipation channel is switched on, e.g., $\kappa/\omega=5\times10^{-6}$~\cite{Pita2024Strong}, the fidelity is $\mathcal{F}=0.995$ for $N=2$, $\mathcal{F}=0.964$ for $N=3$, $\mathcal{F}=0.941$ for $N=4$, $\mathcal{F}=0.912$ for $N=5$, $\mathcal{F}=0.879$ for $N=6$, $\mathcal{F}=0.841$ for $N=7$, and $\mathcal{F}=0.801$ for $N=8$. When the decay rate increases to $\kappa/\omega=2.5\times10^{-5}$, we have $\mathcal{F}=0.982$ for $N=2$, $\mathcal{F}=0.844$ for $N=3$, $\mathcal{F}=0.759$ for $N=4$, $\mathcal{F}=0.668$ for $N=5$, $\mathcal{F}=0.581$ for $N=6$, and $\mathcal{F}=0.502$ for $N=7$. Even when the decay rate is as large as $\kappa/\omega=5\times10^{-5}$, our protocol is still capable of preparing a five-qubit GHZ state with $\mathcal{F}=0.516$. In principle, the running period of our universal control is linear to the system size. In our numerical simulation (on a laptop computer with an Intel Core i5-10400F processor with 2.90 gigahertz in frequency and 32 gigabyte in memory) using the four-order Runge-Kutta method with a fixed yet not optimized time step, the practical time costs $t_c$ are found to be $\sim2.2$ s for $N=2$, $\sim5.7$ s for $N=3$, $\sim24.3$ s for $N=4$, $\sim149.8$ s for $N=5$, $\sim779.1$ s for $N=6$, $\sim3252.0$ s for $N=7$, and $\sim29786.8$ s for $N=8$. Approximately, they fit a polynomial function of $N$ as $t_c\sim\gamma_3N^3+\gamma_2N^2+\gamma_1N+\gamma_0$ with $\gamma_3=0.0716$, $\gamma_2=-0.9006$, $\gamma_1=3.4779$, and $\gamma_0=-4.0368$.

\emph{W state} --- Holding all the interactions among qubits and all the driving fields in the Hamiltonian~(\ref{HamintN}), a single step can be used to generate the $N$-qubit W state. With the driving frequency $\omega_0=-\omega+J$ and strong coupling $J\gg \Omega_n(t)$, the rotating Hamiltonian~(\ref{HameffN1}) with respect to $H_I$ can be rewritten as
\begin{equation}
  H_{\rm eff}^{\rm W}(t)=\sum_{n=0}^{N-1}\Omega_n(t)e^{i\varphi_n(t)}\sigma_{n+1}^+|g\rangle^{\otimes N}\langle g|^{\otimes N}+{\rm H.c.}.
\end{equation}
Considering the $1+N$ model in Fig.~\ref{modelN2}(a) with the maps that $|e_0\rangle\rightarrow|g\rangle^{\otimes N}$ and $|n\rangle\rightarrow\sigma_{n+1}^+|g\rangle^{\otimes N}$, the system can evolve from the ground state $|g\rangle^{\otimes N}$ to the W state $\sum_{n=0}^{N-1}\sigma_{n+1}^+|g\rangle^{\otimes N}/\sqrt{N}$ via one of the passages in Eq.~(\ref{AncillaryN_across}) with the parameters in Eq.~(\ref{ConditionN}) and the proper boundary conditions.

\section{Conclusion}\label{conclusion}

In summary, we develop a full-fledged universal control theory for the popular atomic models with two subspaces, e.g., the assistant and working subspaces. We provide a systematic method for constructing the parametric ancillary basis states crossing the two subspaces and determine the sufficient conditions for them to be useful nonadiabatic passages. In addition, we find the sufficient conditions to convert the static dark states inside the assistant or working subspace to full-featured paths for state engineering. The general theory is applied to prepare the maximal entangled states of distant systems with a high fidelity. By virtue of the longitudinal interaction between neighboring qubits in a superconducting circuit system and local driving fields, we can generate Bell states, $N$-qubit GHZ state, and $N$-qubit W states and realize mutual transformations between the single-excitation Bell state and the double-excitation Bell state. Our work improves the theory framework for nonadiabatic quantum control over discrete systems and is interesting to entangle remote nodes in a future quantum network. It prevails over the existing methods in scalability.

\appendix

\section{A brief recipe of ancillary bases}\label{recipe}

This Appendix contributes to explicitly constructing $M+N$ ancillary bases in Eqs.~(\ref{AncibaseGeneral_high})-(\ref{AncibaseGeneral_across}) and $M+N-2$ bright states in Eq.~(\ref{brightGeneral}) for our general model described in Fig.~\ref{generalmodel} or the full Hamiltonian~(\ref{Hamgeneral}). They follow the orthonormal relations as $\langle\tilde{\mu}_m(t)|\tilde{b}_m(t)\rangle=0$ and $\langle\mu_n(t)|b_n(t)\rangle=0$ with $0\leq m\leq M-2$ and $0\leq n\leq N-2$ and $\langle\mu_{N-1}(t)|\mu_N(t)\rangle=0$. The construction order is suggested as (1) $\{|e_0\rangle, |e_1\rangle\}\rightarrow\{|\tilde{\mu}_0(t)\rangle,|\tilde{b}_0(t)\rangle\}$, $\{|\tilde{b}_0(t)\rangle, |e_2\rangle\}\rightarrow\{|\tilde{\mu}_1(t)\rangle,|\tilde{b}_1(t)\rangle\}$, $\cdots$, $\{|\tilde{b}_{M-3}(t)\rangle, |e_{M-1}\rangle\}\rightarrow\{|\tilde{\mu}_{M-2}(t)\rangle,|\tilde{b}_{M-2}(t)\rangle\}$ for bases in the assistant subspace; (2) $\{|0\rangle, |1\rangle\}\rightarrow\{|\mu_0(t)\rangle,|b_0(t)\rangle\}$, $\{|b_0(t)\rangle, |2\rangle\}\rightarrow\{|\mu_1(t)\rangle,|b_1(t)\rangle\}$, $\cdots$, $\{|b_{N-3}(t)\rangle, |N-1\rangle\}\rightarrow\{|\mu_{N-2}(t)\rangle,|b_{N-2}(t)\rangle\}$ for bases in the working subspace; and eventually (3) $\{|\tilde{b}_{M-2}(t)\rangle,|b_{N-2}\rangle\}\rightarrow|\mu_{N-1}(t)\rangle,|\mu_N(t)\rangle$ for those across the two subspaces.

Using a SU(2)-like rotation matrix, the bases $|\tilde{\mu}_m(t)\rangle$ and $|\tilde{b}_m(t)\rangle$, $0\le m\le M-2$, are determined by the preceding sequence (1) as a parametric superposition over the upper bright state and the upper (assistant) levels. They can be formulated as
\begin{equation}\label{ancibaseAssist}
\begin{aligned}
&\left(\begin{array}{c} |\tilde{\mu}_m(t)\rangle \\ |\tilde{b}_m(t)\rangle \end{array} \right)\\
=&\left(\begin{array}{cc}
\sin\tilde{\theta}_m(t) & \cos\tilde{\theta}_m(t)e^{-i\tilde{\alpha}_m(t)} \\ \cos\tilde{\theta}_m(t) & -\sin\tilde{\theta}_m(t)e^{-i\tilde{\alpha}_m(t)}
\end{array}\right)\left(\begin{array}{c} |\tilde{b}_{m-1}(t)\rangle \\ |e_{m+1}\rangle \end{array} \right),
\end{aligned}
\end{equation}
where $|\tilde{b}_{-1}(t)\rangle\equiv|e_0\rangle$ and $\tilde{\theta}_m(t)$ and $\tilde{\alpha}_m(t)$ are time-dependent coefficients. Similarly, the sequence (2) determines the bases $|\mu_n(t)\rangle$ and $|b_n(t)\rangle$, $0\le n\le N-2$, as a superposition over the down bright state and the down (working) levels:
\begin{equation}\label{ancibaseTarget}
\begin{aligned}
&\left(\begin{array}{c} |\mu_n(t)\rangle \\ |b_n(t)\rangle \end{array} \right)\\
=&\left(\begin{array}{cc}
\cos\theta_n(t) & -\sin\theta_n(t)e^{-i\alpha_n(t)} \\ \sin\theta_n(t) & \cos\theta_n(t)e^{-i\alpha_n(t)}
\end{array}\right)\left(\begin{array}{c} |b_{n-1}(t)\rangle \\ |n+1\rangle \end{array} \right),
\end{aligned}
\end{equation}
where $|b_{-1}(t)\rangle\equiv|0\rangle$ and $\theta_n(t)$ and $\alpha_n(t)$ are time-dependent coefficients. The other two ancillary bases $|\mu_{N-1}(t)\rangle$ and $|\mu_N(t)\rangle$ are then constructed by the last bright states in the two subspaces:
\begin{equation}\label{ancibaseWhole}
\begin{aligned}
&\left(\begin{array}{c} |\mu_{N-1}(t)\rangle \\ |\mu_N(t)\rangle \end{array} \right)\\
&=\left(\begin{array}{cc}
\cos\phi(t) & -\sin\phi(t)e^{-i\alpha(t)} \\ \sin\phi(t) & \cos\phi(t)e^{-i\alpha(t)}
\end{array}\right)\left(\begin{array}{c} |b_{N-1}(t)\rangle \\ |\tilde{b}_{M-2}(t)\rangle \end{array} \right),
\end{aligned}
\end{equation}
where $\phi(t)$ and $\alpha(t)$ are time-dependent coefficients. Finally, we obtain a completed and orthonormal set of bases to span the whole Hilbert space of an $M+N$-dimensional system.

\bibliographystyle{apsrevlong}
\bibliography{ref}

\end{document}